\documentclass[12pt]{msml2020} 

\title[Schr\"odingeRNN]{Schr\"odingeRNN: Generative Modeling of Raw Audio as a Continuously Observed Quantum State}
\usepackage{times}
\usepackage{braket}
\usepackage{comment}
\usepackage{dsfont}
\usepackage{amsfonts}
\usepackage{bm}

\msmlauthor{%
\Name{Be\~nat {Mencia Uranga}} \Email{beinat.mencia@gmail.com}\\
\and
 \Name{Austen Lamacraft} \Email{al200@cam.ac.uk}\\
 \addr TCM Group, Cavendish Laboratory, University of Cambridge, J. J. Thomson Ave., Cambridge CB3 0HE, UK%
}

\newcommand*{\1}{\mathds{1}}

\newcommand*{\bx}{\mathbf{x}}

\newcommand*{\bs}{\mathbf{s}}

\newcommand*{\bq}{\mathbf{q}}
\newcommand*{\bg}{\mathbf{g}}

\newcommand*{\bbf}{\mathbf{f}}

\newcommand*{\bA}{\mathbf{A}}
  
\newcommand*{\bP}{\mathbf{P}}  

\newcommand*{\bC}{\mathbf{C}}
\newcommand*{\bF}{\mathbf{F}}
\newcommand*{\bQ}{\mathbf{Q}}     
\newcommand*{\bH}{\mathbf{H}}

\newcommand*{\bL}{\mathbf{L}}

\newcommand*{\cH}{{\mathcal H}}

\newcommand*{\cP}{{\mathcal P}}
\newcommand*{\cO}{{\mathcal O}}

\newcommand*{\tC}{{\text{C}}}

\newcommand*{\tbm}{{\textbf{m}}}
\newcommand*{\w}{{\omega}}

\DeclareMathOperator{\E}{\mathbb{E}}
\DeclareMathOperator{\tr}{tr}

\newcommand{\subfigimg}[3][,]{%
	\setbox1=\hbox{\includegraphics[#1]{#3}}
	\leavevmode\rlap{\usebox1}
	\rlap{\hspace*{0pt}\raisebox{\dimexpr\ht1+0\baselineskip}{#2}}
	\phantom{\usebox1}
}

\begin{document}

\maketitle

\begin{abstract}%
  We introduce Schr\"odingeRNN, a quantum inspired generative model for raw audio. Audio data is wave-like and is sampled from a continuous signal. Although generative modelling of raw audio has made great strides lately, relational inductive biases relevant to these two characteristics are mostly absent from models explored to date.
  Quantum Mechanics is a natural source of probabilistic models of wave behaviour. Our model takes the form of a stochastic Schr\"odinger equation describing the continuous time measurement of a quantum system, and is equivalent to the \textit{continuous Matrix Product State} (cMPS) representation of wavefunctions in one dimensional many-body systems. This constitutes a deep autoregressive architecture in which the system€™s state is a latent representation of the past observations. We test our model on synthetic data sets of stationary and non-stationary signals. This is the first time cMPS are used in machine learning.%
\end{abstract}

\begin{keywords}%
  Machine Learning, Generative Models, Quantum Physics, Matrix Product States.%
\end{keywords}

\section{Introduction}

Audio generation appears in different machine learning tasks such as music synthesis or \textit{text-to-speech}, where the input is text and the output is speech audio. One of the reasons why it is challenging is that the dimensionality of the raw audio signal is usually a lot larger than that of the effective semantic-level signal. In speech synthesis for instance, one is typically interested in generating utterances corresponding to full sentences. At a minimum quality sampling rate of 16kHz, an average of 6,000 samples per word are generated \cite{Mehri:2017}.

Both music and speech are complex and highly structured. In audio signal form, different features have different timescales, ranging from milliseconds to minutes in the case of music. Because the correlations span different orders of magnitude, modeling the temporal correlations of the signals is challenging  \cite{Dieleman:2018}.

Traditionally, the high dimensionality of the raw audio modelling problem has been dealt with by compressing the audio waveforms into spectral or higher level features, and then defining generative models on these features. Examples in music generation are symbolic representations such as scores and MIDI sequences. The compression causes many of the subtleties that are crucial for the quality of sound to vanish. A way around these limitations is to model sound in the raw audio domain instead. While the digital form of audio is also lossy, the relevant information for the quality of musicality is retained.

There has been recent work on raw audio modelling using autoregressive models: AMAE \cite{Dieleman:2018}, WaveNet \cite{Oord:2016}, VRNN \cite{Chung:2016}, WaveRNN \cite{Kalchbrenner:2018} and SampleRNN \cite{Mehri:2017}. The first is a convolutional neural network with dilated convolutions, the rest are recurrent neural networks. Beyond autoregressive models, there is WaveGlow \cite{Prenger:2018} where a flow model is used and WaveGAN \cite{Donahue:2018} using generative adversarial networks.

\section{Quantum inspired machine learning}

A natural connection between quantum mechanics and machine learning is that probability distributions appear in both disciplines. Quantum inspired machine learning is the use of quantum wave functions and quantum processes to model probability distributions and generative processes. In each case, one needs to choose the wave function and the physical process that is suitable for the problem at hand.

In raw audio modeling, the data is wave-like and quantum mechanics is a natural source of probabilistic models of wave behaviour. Hence, quantum inspired models might benefit from the \textit{inductive bias} induced by these two characteristics: the wave-like and probabilistic nature. Furthermore, within the range of problems that exist in machine learning, one-dimensional machine learning is specially appealing for quantum many-body physicists. This is because in physics, the most powerful numerical and analytical tools have been developed to study one-dimensional systems. Therefore, there is the potential to use them to solve machine learning tasks. In this work, we use \textit{continuous matrix product states} (cMPS) (see Appendix \ref{appendix:Continuous matrix product states}), a numerical tool used in many-body quantum physics to deal with Hilbert spaces of many-body systems, to handle the high dimensionality of the audio data. There has been several works where \textit{matrix product states} and more general tensor networks have been used for machine learning \cite{Glasser:2019, Cheng:2019, Stokes:2019, Li:2018, Han:2017, Efthymiou:2019, Liu:2019, Evenbly:2019, Guo:2018, Glasser:2018, Stoudenmire:2018, Novikov:2016, Bradley:2019, Stoudenmire:2016}. As far as we are aware, this is the first work where cMPS are  used. The code is available in \cite{audiomps}.

\section{A quantum-inspired model for raw audio}

In a typical raw audio dataset, each data point is a vector with  several thousands of real valued elements, e.g.  in NSynth dataset \cite{Engel:2017} each note amounts to 64,000 samples. Hence, data lives in a very high dimensional space, which makes it unaffordable to explore brute-force: we are faced with \textit{the curse of dimensionality}. 

This is reminiscent of a problem that arises in many-body quantum optimization problems. When trying to find the variational ground state of a many-body quantum system, one has an exponentially large Hilbert space to explore. Matrix product states (MPS) serve as a tool to overcome the curse of dimensionality in this context. As explained in \cite{Orus:2014}, it gives a way to parameterize the relevant corner of the Hilbert space efficiently.

 The fact that MPS has proven to be a successful tool to overcome the curse of dimensionality in physics suggests that it might be useful in machine learning as well. In this work, we want to explore the utility of MPS to model raw audio. On the other hand, MPS is not suitable for modeling continuous data (like raw audio), because it describes lattices of discrete degrees of freedom like spins. As explained in Appendix~\ref{appendix:Continuous matrix product states}, there exists a generalization of MPS to systems with continuous degrees of freedom: \textit{continuous matrix product states} (cMPS).
 
 We will be thinking of the audio waveforms as the outcome of a sequential measurement of a continuous observable throughout the evolution of a quantum system.

 \subsection{The Schr\"odingeRNN model}
 
 As explained in Appendix \ref{apendix:Physical picture of cMPS}, our model generative process consists on the continuous measurement of the homodyne current $I_t$ (see Appendix \ref{appendix:Balanced homodyne detection}), on the output of an open quantum system described by a cMPS. As a refinement of the cMPS model, we include two extra learning variables: $A$ and $\sigma$.The model involves the signal $I_t$ together with a latent Hilbert space consisting of vectors $\ket{\psi}\in\mathbb{C}^D$. The signal follows the stochastic process

\begin{equation}\label{It}
I_{t+dt} = A\langle R_t+R_{t}^\dagger \rangle_t + z, \;\; \text{where} \;\; z \sim N(0,1/dt).
\end{equation}
The parameter $A$ is a real learning variable, $R_t= e^{iHt}Re^{-iHt}$ ($H$ is real and diagonal), $R\in \mathbb{C}^{D\times D}$  is a matrix acting on the latent space and as before the angular brackets $\langle\cdot\rangle_t$ denote the quantum mechanical expectation over an (unnormalized) state $\ket{\tilde\psi_t}$

\begin{equation}
\langle \cdot\rangle_t = \frac{\braket{\tilde\psi_t|\cdot|\tilde\psi_t}}{\braket{\tilde\psi_t|\tilde\psi_t}}.
\end{equation}
The state $\ket{\tilde\psi}$ evolves according to 

\begin{align}\label{SRNN: EOM}
\ket{\tilde\psi_{t+dt}} &= \left[\1 - \frac{\sigma^2}{2}R_{t}^\dagger R_{t}  dt+R_{t} I_{t+dt} dt\right]\ket{\tilde \psi_t}, \\
\label{SRNN: EOM_1}
\ket{\psi_{t+dt}} &= \ket{\tilde\psi_{t+dt}} / \sqrt{\langle \tilde\psi_{t+dt} | \tilde\psi_{t+dt} \rangle}.
\end{align}
The purpose of introducing the training variable $A$ is to learn the amplitude of the signal. The amplitude is set by $A\langle R_t+R_{t}^\dagger \rangle_t dt$ in Eq.~\eqref{SDE_model}. 
This is done to learn $R$ independently of the amplitude of the signals in the dataset. This way, 
the training of $R$ is geared solely towards optimizing the time evolution of $\ket{\tilde\psi_t}$ in Eq.~\eqref{SSE}. The hyperparameter $dt$ sets the strength of the term $R_{t}^{\dagger}R_t$ compared to $R_{t} I_{t+dt} dt$. In cases where we are interested in fixing $dt$ to be the real time discretization of the data (see Sec. \ref{Learnable parameters and hyperparameters}), is $\sigma$ the hyperparameter in charge of this. The initial state $\ket{\psi_0}$ is learned.

The conditional joint probability density for a sequence of measurements $\{ I_t\}$ is

\begin{align}\label{autoregressive}
\nonumber
p(I_T, ..., I_1 |H,R,A, \ket{\psi_0}) &= \prod_{k=0}^{T-1} p(I_{k+1}| I_k, ...,I_1; H,R,A, \ket{\psi_0}) , \text{ where} \\
p(I_{k+1}| I_k, ...,I_1; H,R,A, \ket{\psi_0})  &=   \sqrt{\frac{dt}{2\pi}}\exp\left[-\frac{dt}{2} \left (I_{k+1} - \langle R_k + R_{k}^\dagger \rangle_k \right )^2 \right].
\end{align}
This constitutes an \textit{Autoregressive Recurrent Neural Network} where the hidden state is the quantum wavefunction $\ket{\tilde\psi}$ and the non-linear update equation is Eq.~\eqref{SRNN: EOM_1}.
Aside from a few last details that will be explained in the coming sections, this probability distribution defines our \textit{quantum-inspired model}. A detailed derivation of the model can be found in Appendix \ref{apendix:Physical picture of cMPS}.

\section{Data as homodyne current}

We have seen that the sequential measurement of the homodyne current on the output of an open quantum system described by a cMPS gives rise to the autoregressive probability distribution shown in Eq.~\eqref{autoregressive}. We now want to use this probability distribution to model raw audio data $x_t$. One obvious thing to do is to consider the raw audio to be the homodyne current, i.e. $I_t \equiv x_t$, in which case the generative model is defined as

\begin{align}\label{px}
\nonumber
p(x_T, ..., x_1 | H,R,A, \ket{\psi_0})  &= \prod_{k=0}^{T-1} p(x_{k+1}| x_k, ...,x_1;  H,R,A, \ket{\psi_0}) , \\
p(x_{k+1}| x_k, ...,x_1;  H,R,A, \ket{\psi_0})  &=   \sqrt{\frac{dt}{2\pi}}\exp\left[-\frac{dt}{2}  \left (x_{k+1} - \langle  R_k + R_{k}^\dagger \rangle_k \right )^2  \right].
\end{align}
Note that in the limit $dt \rightarrow 0$, the variance $1 /dt$ diverges and the samples of this probability distribution become pure noise. Hence, this model does not have a continuous limit in the sense that as the time discretization becomes dense, the signal does not become smoother but more discontinuous.

If our training strategy is \textit{maximum log likelihood}, the loss function of a single data point is

\begin{equation}
 - \log p(x_T, ..., x_1 |H,R,A, \ket{\psi_0}) = -\sum _{k=0}^{T-1} \log p(x_{k+1}| x_k, ...,x_1; H,R,A, \ket{\psi_0}).
\end{equation}
Since $dt$ is a hyperparameter (i.e., we do not learn it), we define the loss function as

\begin{equation}\label{loss_I_eq_x}
\text{loss} (H,R,A, \ket{\psi_0}) = \sum _{k=0}^{T-1} \left (x_{k+1} - \langle  R +  R^\dagger \rangle_k \right )^2.
\end{equation}
At sampling time, the variance of the Gaussian in Eq.~\eqref{px} is tuned by introducing a temperature parameter $T$ (explained in Sec.~\ref{sampling}) to optimize the quality of the samples. Therefore, $dt$ does not influence the variance at generation time.

\section[Stochastic differential equation perspective]{Time derivative of data as homodyne current: a \\
stochastic differential equation perspective}
\label{sec:SDE}

Let us consider Eq.~\eqref{It}. Note that $z \sim N(0, q^2)$ is equivalent to $q z \sim N(0, 1)$ and therefore multiplying both sides by $dt$

\begin{equation}\label{I_dt}
I_{t+dt} dt =\langle R_t +  R_{t}^\dagger \rangle_t dt + d\beta_t.
\end{equation}
The process $\beta_t$ is Brownian motion and its independent increments have variance $dt$. In the limit $dt \rightarrow 0$, this equation is reminiscent of a \textit{stochastic differential equation} 

\begin{equation}\label{SDE_inc}
 dI_t =\text{f}(I_t,t) dt + d\beta_t.
\end{equation}
On the other hand, the left hand side of Eq.~\eqref{I_dt} contains the value of the stochastic process $I_t$ whereas the left hand side of Eq.~\eqref{SDE_inc} contains the differential of the process $dI_t \equiv I_{t+dt} - I_t$. In order to rephrase our model in the language of stochastic differential equations, an option is to define the time derivative of the raw audio data to be the outcome of the homodyne current measurement, i.e. $I_t \equiv dx_t /dt$, instead of $I_t \equiv x_t$. 

As explained in \cite{Chen:2018}, there are several advantages of having a continuous formulation of the model, even though one always needs to discretize to perform numerical calculations. One of the main advantages is that one can use the machinery developed to numerically integrate stochastic differential equations.

Even though it is appealing to rephrase the model in terms of SDEs, later we will see that this is not always a good option, since for certain datasets, the time derivative $d x_t /dt$ of the signals is more spiky and discontinuous than the signal $x_t$, which is problematic for training our model. In the remainder of the chapter, we will use both approaches. Depending on the choice, the notation will be 

\begin{itemize}
  \item $\text{If }I_t \equiv x_t, \text{ then}   \ x_{t+dt}  =\langle R_t + R_{t}^\dagger \rangle_{t} + z, \text{ where } z  \sim N(0,1/dt).$
  \item $\text{If }I_t \equiv d x_t /dt, \text{ then}  \ dx_t =\langle R_t + R_{t}^\dagger \rangle_{t}dt + d\beta_t$.
\end{itemize}
Throughout the paper, unless we consider it helpful, we do not specify the units of different quantities.

\subsection{The model from an SDE perspective}

In the continuous formulation of the model, the signal follows the stochastic process (It\^o process)

\begin{equation}
\label{SDE_model}
dx_t = A\langle R_t+R_{t}^\dagger \rangle_t dt + d\beta_t.
\end{equation}
Here $\beta_t$ is Brownian motion with diffusion constant $q$, which means that the independent increments $\Delta \beta \equiv \beta_{k+1}-\beta_k$ are zero mean Gaussian random variables with variance $q \Delta t$. The state $\ket{\tilde\psi}$ evolves according to 

\begin{equation}
\label{SSE}
d\ket{\tilde\psi_t} = \left[-\frac{\sigma^2}{2}R_{t}^\dagger R_t  dt+R_t dx_t\right]\ket{\tilde \psi_t}.
\end{equation}
Hence our model in Eq.~\eqref{SDE_model} has the form of a non-linear stochastic differential equation.

\subsection{Generalization to density matrices}

 To add expressivity to the model, we can consider starting from a learned density matrix $\rho_0$, and evolving the density matrix instead of the state $\ket{\tilde\psi_t}$. There is the disadvantage that $\rho\in \mathbb{C}^{D\times D}$, so it is more costly to evolve than the pure state. The equation of motion for the (unnormalized) density matrix is

\begin{equation}\label{SSE_rho}
\frac{d\tilde\rho_t}{dt} =  \sigma^2 L (\tilde\rho_t)  + (\tilde\rho_t R_{t}^{\dagger}+R_t \tilde\rho_t) \frac{dx_t}{dt},
\end{equation}
where $L (\cdot)$ is the Linbladian

\begin{equation}
L (\rho) = R_t \rho R_{t}^{\dagger} - \frac{1}{2} \left ( R_{t}^{\dagger}R_t \rho + \rho R_{t}^{\dagger}R_t \right ).
\end{equation}
The quantum mechanical average in Eq.~\eqref{SDE_model} then becomes

\begin{equation}
\langle R_{t} + R_{t}^{\dagger} \rangle_t = \frac{\text{Tr} \left [ \left ( R_t + R_{t}^{\dagger} \right ) \tilde\rho_t \right ]}{\text{Tr} \left [ \tilde\rho_t \right ]}.
\end{equation}

\subsection{Parameter estimation}
\label{Parameter estimation}

We now have a parametric form of our model and we need to find the values of the parameters that best fit the data, given a dataset. The probability distribution of continuous processes is not normalizable \cite{Sarkka:2019}, i.e. if we formally define it as

\begin{equation}
p(\mathcal{X}_t)= \lim_{n\to\infty}  p(x_1, ..., x_n),
\end{equation}
this limit tends to zero or infinity almost everywhere in the domain of the distribution. In order to define a finite loss function, we can consider the relative probability distribution of the process $\mathcal{X}_t$ with respect to the probability of another process that does not contain the learnt parameters. It is natural to define the relative probability of the signal $\mathcal{X}_t$ with respect to the driving Brownian motion $\beta_t$. Let us call the probability measure of our model $\mathbb{P}_{\text{cMPS}}(\mathcal{X}_t)$ and the probability distribution associated with Brownian motion $\mathbb{P}_{\beta}(\mathcal{X}_t)$. According to the Girsanov theorem (\cite{Sarkka:2019}), the relative probability of $\mathbb{P}_{\text{cMPS}}(\mathcal{X}_t)$ with respect to $\mathbb{P}_{\beta}(\mathcal{X}_t)$ is given by the Radon-Nikodym derivative involved in changing measure from $\mathbb{P}_{\text{cMPS}}(\mathcal{X}_t)$ to $\mathbb{P}_{\beta}(\mathcal{X}_t)$:

\begin{equation}\label{RN}
\frac{d\mathbb{P}_{\text{cMPS}}(\mathcal{X}_t)}{d\mathbb{P}_{\beta}(\mathcal{X}_t)}  = \exp\left(\frac{A}{q}\int\langle R_t+R_{t}^\dagger\rangle_t dx_t - \frac{A^2
}{2 q}\int\langle R_t+R_{t}^\dagger\rangle_t^2 dt\right),
\end{equation}
where  $q$ is the diffusion constant of the Brownian motion. Our training strategy is to minimise the negative log likelihood (relative to the measure of Brownian motion), i.e., our loss function (associated to a single continuous audio signal $\mathcal{X}_t$) is minus the logarithm of Eq. \eqref{RN}:

\begin{equation}\label{loss_continuous}
\text{loss} = -\log \frac{d \mathbb{P}_{\text{cMPS}}(\mathcal{X}_t)}{d \mathbb{P}_{\beta}(\mathcal{X}_t)} = -A\int\langle R_t+R_{t}^\dagger\rangle_t dx_t + \frac{A^2
}{2 }\int\langle R_t+R_{t}^\dagger\rangle_t^2 dt.
\end{equation}
Note that we removed the factor $1/q$ because it is not a learning variable. 

\subsection{Discretization}

Since we cannot solve Eq.~\eqref{SDE_model} exactly, we cannot evaluate the likelihood exactly and so we need the aid of discretization methods. We use the Euler-Maruyama integration scheme

\begin{align}
\ket{\tilde\psi_{t+\Delta t}} &= \left[\1-\frac{\sigma^2}{2}R_{t}^\dagger R_t\Delta t+R_t \Delta x_t\right]\ket{\tilde \psi_t}, \;\; \text{and} \\
\tilde\rho_{t+\Delta t}&= \left[\1-\frac{\sigma^2}{2}R_{t}^\dagger R_t\Delta t+R_t \Delta x_t\right] \tilde \rho_t
\left[\1-\frac{\sigma^2}{2}R_{t}^\dagger R_t\Delta t+R_t \Delta x_t\right]^\dagger.
\end{align}
The discretization of the model SDE \eqref{SDE_model} is

\begin{equation}
\Delta x_t = A\langle R_t+R_{t}^\dagger \rangle_t \Delta t + \Delta \beta_t,
\end{equation}
where $\Delta x_t \equiv x_{t+\Delta t}-x_t$  and $\Delta \beta_t \equiv \beta_{t+\Delta t}-\beta_t$. The discretization of the loss function \eqref{loss_continuous} is 

\begin{equation}
\text{loss} (H,R,A, \ket{\psi_0})  = -A \sum_t \langle R_t+R_{t}^\dagger\rangle_t \Delta x_t +
 \frac{A^2}{2} \sum_t \langle R_t+R_{t}^\dagger\rangle_t^2 \Delta t.
\end{equation}
Neglecting the constant $-{\Delta x_t}^2/2 \Delta t$ and the multiplicative factor $\Delta t /2$, it is expressed as

\begin{equation}
\text{loss} (H,R,A, \ket{\psi_0}) = \sum_t  \left ( \frac{\Delta x_t}{\Delta t} - A \langle R_t+R_{t}^\dagger\rangle_t  \right )^2.
\end{equation}
Note that substituting $\Delta x_t / \Delta t$ by $x_t$, this is equal to Eq.~\eqref{loss_I_eq_x}.

\subsection{Learnable parameters and hyperparameters}
\label{Learnable parameters and hyperparameters}

The model is specified by the learnable parameters $A,H,R$ and $\ket{\psi_0}$ (or $\rho_0$) and the hyperparameters:

\begin{enumerate}
	\item The bond dimension $D$, which reflects the complexity of the model.
	
	\item Time discretization $\Delta t$. If we set it to be equal to the inverse of the sampling rate of the data, which corresponds to matching the time discretization of the data with the time discretization of the model, we can relate the eigenvalues $H$ to the (angular) frequencies $\w = 2 \pi f$ of the data (see Appendix \ref{appendix:regularization}).
		
	\item $\sigma$, which governs the strength of the term $R^{\dagger}R$.
	
	\item The hyperparameters $\sigma_{\omega}$ and $\sigma_R$ are regularisers for $H$ and $R$ (see Appendix \ref{appendix:regularization}). In this work we do not experiment with regularisers.
	
\end{enumerate}

\subsection{Sampling}
\label{sampling}

After training, we use the learnt parameters $H,R,A$ and $\ket{\psi_0}$ (or $\rho_0$) to generate samples using the discrete model

\begin{align}
\nonumber
x_{t+1} &= x_t + \Delta x_t, \text{ where,} \\
\Delta x_t &= A\langle R_t+R_{t}^\dagger \rangle_t \Delta t + \sqrt{T} \Delta \beta_t.
\end{align}
We introduce a temperature parameter $T$ to tune the variance of the independent increments of the Brownian motion. In generative modeling, it is common to introduce a temperature parameter to optimize the quality of the sampling. See for example \cite{Kingma:2018}. At $T=0$ the generative process is deterministic. As the temperature is increased, the generative model gives rise to a variety of samples due to the randomness of the increments. At very high temperatures, the Gaussian noise dominates the generative process and the samples resemble the training data less and less.

\section{Experiments}
\label{results}

To test the capabilities of our model, we create synthetic datasets where we know the ground truth probability distributions. We can then readily check whether the learnt probability distribution matches the ground truth. We train on three different datasets: damped sines with random delays, \textit{Gaussian processes} and \textit{filtered Poisson processes}.

\subsection{Damped sines}
\label{damped_sines}

The experimental details are shown in Appendix \ref{appendix: damped sines}.

\subsubsection{Single frequency experiment}
\label{1_freq_exp}

\begin{figure}[h]
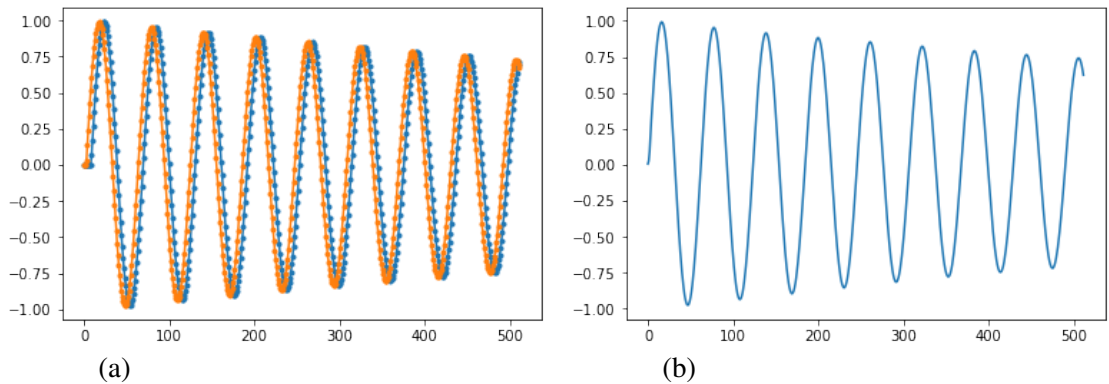

\centering
\subfigimg[width=7.4cm,angle=0]{\raisebox{-145pt}{\hspace*{35pt} (a)}}{1a} 
\subfigimg[width=7.4cm,angle=0]{\raisebox{-145pt}{\hspace*{35pt} (b)}}{1b}
\caption{(a) Two signals of the dataset with different delays. The length of the data samples is 512 and the sampling frequency 16 kHz. (b) The $T=0$ sample from the our pure state model, after training. It has the form of a damped sine with a finite delay.}
\label{1freq_a}
\end{figure}

We start by modeling a dataset that consists of damped sines with random delays. Each signal has amplitude zero at the beginning, and the length of this ``silence'' period is random (see two samples in Fig.~\ref{1freq_a}(a)). All signals have the same frequency $f = 261.6$Hz, the sampling rate is 16KHz and the length of each signal vector is 512 (which corresponds to 0.032 seconds). To generate the training set, we obtain the random delays by sampling from the distribution Gamma$(\alpha=2, \beta=0.39)$.  

\begin{figure}[h]
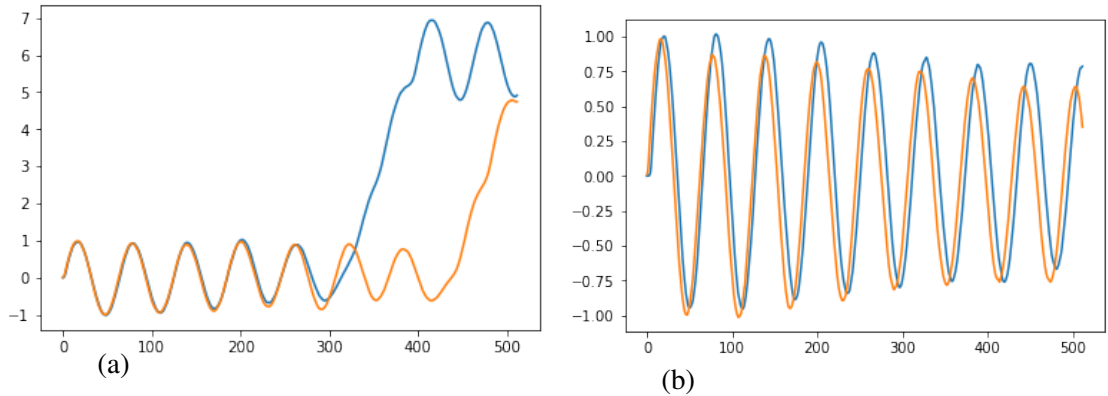

\centering
\subfigimg[width=7.4cm,angle=0]{\raisebox{-145pt}{\hspace*{35pt} (a)}}{2a} 
\subfigimg[width=7.4cm,angle=0]{\raisebox{-145pt}{\hspace*{35pt} (b)}}{2b}
\caption{(a) Two samples at $T=30$, $D=100$ and $\sigma = 10^{-4}$ using the pure state model. The shape of a damped sine is well captured. On the other hand, we can only get proper samples of length 300 approximately. (b) Two samples at $T=42$, $D=100$ and $\sigma = 10^{-4}$, where we use a density matrix. Unlike in the pure state case, the samples look like damped sines, for the whole length of 512 samples.}
\label{samp_1freq}
\end{figure}

We start by considering the pure state model. The results are:

\begin{enumerate}

 \item At $T=0$, the sampling is deterministic given the learned initial state $\ket{\psi_0}$, as explained in Sec.~\ref{sampling}. The zero temperature sample has the shape of a damped sine with a finite delay. This sample is shown in Fig.~\ref{1freq_a}(b).  
 \item At finite temperature, we find that we can capture the delay degree of freedom, i.e. different samples have the form of a damped sine, with different delays. On the other hand, we find that the samples have the right form (i.e., the form of a damped sine) for the first 300 points only, having been trained on signals of length 512. In this sense, the outcome of this experiment is not very satisfactory. We experimented with different bond dimensions up to $D=300$. We show two samples in Fig.~\ref{samp_1freq}(a).
 \end{enumerate}

We now consider the time evolution of a density matrix 

\begin{align}
\tilde\rho_{t+\Delta t}&= \left[\1-\frac{\sigma^2}{2}R^\dagger(t) R(t)\Delta t+R(t) \Delta x_t\right] \tilde \rho_t
\left[\1-\frac{\sigma^2}{2}R^\dagger(t) R(t)\Delta t+R(t) \Delta x_t\right]^\dagger.
\end{align}
In this case, sampling remains of good quality up to 512 samples, as can be seen in Fig.~\ref{samp_1freq}(b).

We also experiment with damped sines of two different frequencies. We find that the model learns the manifold of damped sines fairly well, but it fails to capture the two frequencies degree of freedom of the dataset. Details about this experiment are shown in Appendix \ref{appendix: Two frequencies experiment}.

\subsection{Gaussian processes}
\label{GP}

In the previous section, we tested  the ability of our model to learn damped sines. On the other hand, real life sound is a lot more complex than sine waves. For example, real sound is made of several harmonics (unlike a sine wave). To test the capabilities of our model on more realistic data, we move on to training on \textit{Gaussian processes}, specifically \textit{Mat\'ern spectral mixtures} (see Appendix \ref{appendix: details about GP}). The experimental details are shown in Appendix \ref{appendix: GP and FPP}.

\begin{figure}[h]
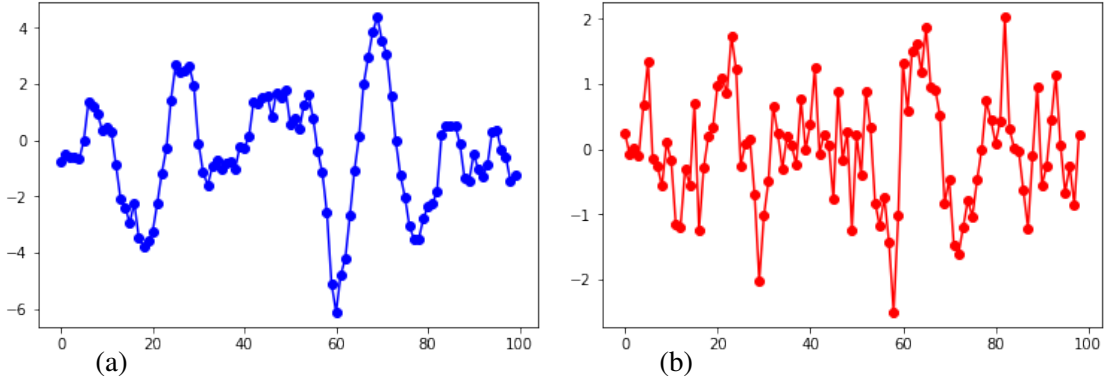

\centering
\subfigimg[width=7.4cm,angle=0]{\raisebox{-145pt}{\hspace*{35pt} (a)}}{3a} 
\subfigimg[width=7.4cm,angle=0]{\raisebox{-145pt}{\hspace*{35pt} (b)}}{3b}
\caption{(a) A sample $x_t$ of a Gaussian process with \textit{Mat\'{e}rn spectral mixture} spectral function defined by $(\sigma, \lambda, \w_0)=(2, 50, 300)$. (b) The increments $\Delta x_t = x_{t+dt}-x_t$ of the sample shown in (a).}
\label{x_vs_dx}
\end{figure}

\begin{figure}[h]
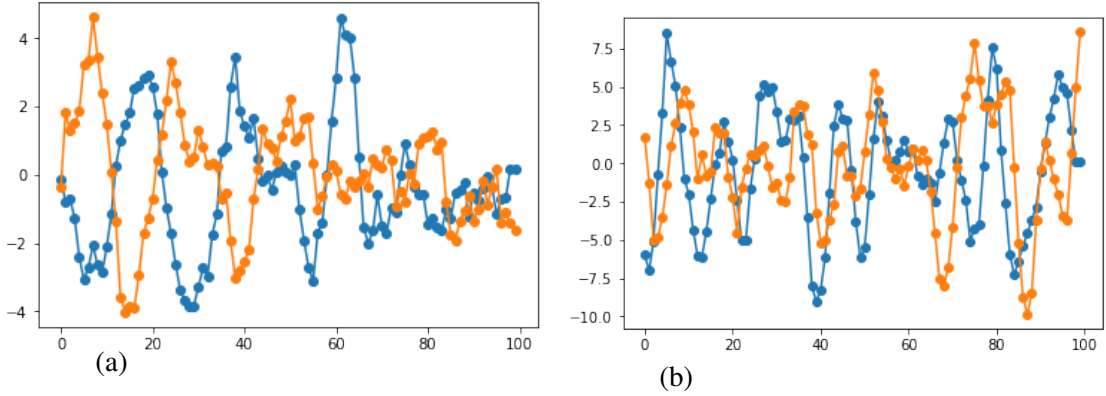

\centering
\subfigimg[width=7.4cm,angle=0]{\raisebox{-145pt}{\hspace*{35pt} (a)}}{4a} 
\subfigimg[width=7.4cm,angle=0]{\raisebox{-145pt}{\hspace*{35pt} (b)}}{4b}
\caption{Samples from the two dataset we train on. Two Gaussian processes with \textit{Mat\'{e}rn spectral mixture} spectral function defined by (a) $(\sigma', \lambda, \w_0)=(2, 50, 300)$ and (b) $(\sigma'_i, \lambda_i)=(2, 50)$ for $i=1,2,3$ and $(\w_1,\w_2,\w_3)=(300,500,700)$.}
\label{Gaussian_data}
\end{figure}

We create a dataset of samples of a stationary Gaussian process of choice. We generate the data using a discrete stochastic equation (see Appendix.~\ref{equivalent_disc}) instead of sampling from a multivariate Gaussian distribution. 
 We train on two different \textit{Mat\'{e}rn spectral mixture} processes. In the first, the spectral function consists of a single pair of Lorentzians centered at $\w_0 = \pm 300$ and  $(\sigma, \lambda)=(2, 50)$. In the second, we consider a mixture of three frequencies. The mixture is defined by the parameters $(\sigma_i, \lambda_i)=(2, 50)$ for $i=1,2,3$ and $(\w_1,\w_2,\w_3)=(300,500,700)$. We show two samples from each of the two datasets in Fig.~\ref{Gaussian_data}.

\subsubsection{Results}

\begin{figure}[h]
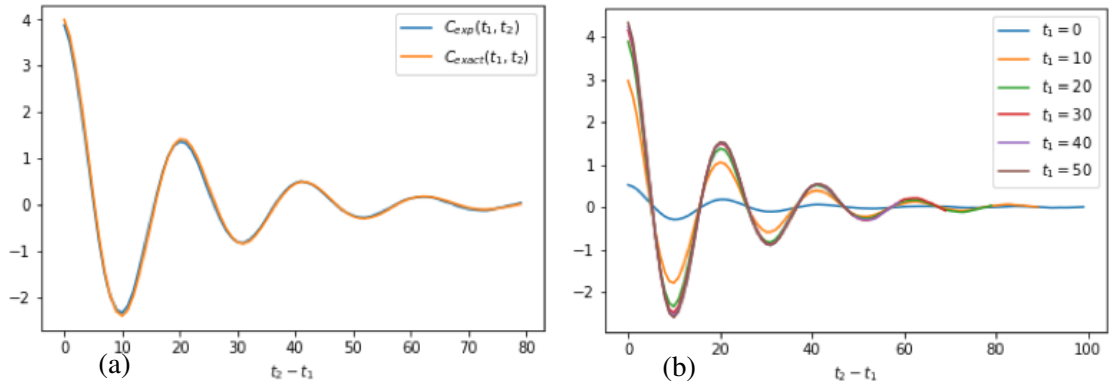

\centering
\subfigimg[width=7.4cm,angle=0]{\raisebox{-145pt}{\hspace*{35pt} (a)}}{5a}
\subfigimg[width=7.4cm,angle=0]{\raisebox{-145pt}{\hspace*{35pt} (b)}}{5b} 
\caption{The time indicated in the horizontal axis is an integer index that specifies the time step of the discrete covariance. (a) Perfect match of experimental (blue) and exact (yellow) covariances. The exact covariance has parameters $(\sigma', \lambda, \w)=(2, 50, 300)$. The experimental covariance is calculated using 40000 samples at $T=0.00051$. The hyperparameters used are $D=50$, $dt=0.001$ and $\sigma=1$. (b) Experimental covariance $\tC_{\text{exp}}(t_1,t_2)$ for different initial times, showing stationarity. It reaches stationarity at $t_1 \approx 20$.}
\label{Dmix1}
\end{figure}

Due to the higher complexity of the data compared to the damped sines in Sec.~\ref{damped_sines}, instead of just looking at plots of samples, we judge whether the model is successful at learning the above process by 1) calculating the experimental covariance from $N$ samples

\begin{equation}
C_{\text{exp}} (t, t') \equiv \frac{1}{N} \sum_{i=1}^{N} x_i (t) x_i (t'),
\end{equation}
and comparing it with the exact covariance and 2) checking that the experimental covariance is stationary, i.e. $C_{\text{exp}}(t,t')=C_{\text{exp}}(\tau)$. We find that the model is successful at learning this process and we show the results on Fig.~\ref{Dmix1}. On the other hand, as explained in Sec.~\ref{sampling}, sampling depends on temperature. The experimental covariance only matches the exact covariance at a given temperature. As one departs from this temperature, the two covariances start to differ. Similarly, the samples are stationary only in a small range of temperatures around this temperature. Furthermore, we find that the experimental covariance becomes stationary only after a few steps, not from the beginning.

%

\begin{figure}[h]
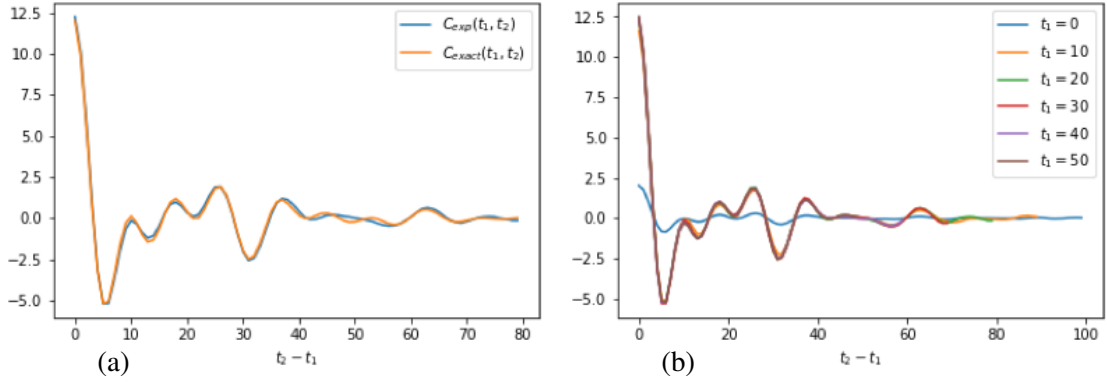

\centering
\subfigimg[width=7.4cm,angle=0]{\raisebox{-145pt}{\hspace*{35pt} (a)}}{6a}
\subfigimg[width=7.4cm,angle=0]{\raisebox{-145pt}{\hspace*{35pt} (b)}}{6b} 
\caption{The time indicated in the horizontal axis is an integer index that specifies the time step of the discrete covariance. (a) Perfect match of experimental (blue) and exact (yellow) covariances.  The exact covariance has parameters $(\sigma'_i, \lambda_i)=(2, 50)$ for $i=1,2,3$ and $(\w_1,\w_2,\w_3)=(300,500,700)$. The experimental covariance is calculated using 40000 samples at $T=0.002$. The hyperparameters used are $D=100$ and $dt=0.001$. (b) Experimental covariance $\tC_{\text{exp}}(t_1,t_2)$ for different initial times, showing stationarity. Before $t_1 = 10$ the experimental covariance is non-stationary.}
\label{Dmix3}
\end{figure}

When trained on a mixture of three frequencies, the model succeeds at reproducing stationary samples (after a given time) with the right covariance function. The results are shown in Fig.~\eqref{Dmix3}.

\subsection{Poisson processes}

A feature of stationary Gaussian processes is that because the covariance function is symmetric $\tC(t,t')=\tC(t',t)$ and all diagonal elements are equal to $\tC(t,t)$, the probability density of a given sample $x(t)$ is the same as the probability density of the time-inverted sample. We refer to this symmetry as \textit{time-reversal symmetry} (TRS). 

Many real life sounds are not time-reversal symmetric. For example, the chirp of a bird will sound different if played backward. Therefore time-reversal symmetric models like Gaussian models are not suitable to model this kind of sound.

Our cMPS based model is not constrained by time-reversal symmetry, as multivariate Gaussian probability distributions are. We can see this by looking at the discretized time evolution of the unnormalized state. The fact that the one-step time evolution operator does not commute with itself at different times, implies the absence of the TRS constraint.

One can check whether a probability distribution is time-reversal symmetric, from certain correlation functions. Consider the two correlators

\begin{align}\label{TRS_correlators}
\nonumber
\E \left [ x^3(t_i) x(t_j) \right ] &= \int d x(t_1)...dx(t_N)  \ x^3(t_i) x(t_j) \ p \left (x(t_1),...,x(t_i),...,x(t_j),...,...,x(t_N) \right ), \\
\E \left [ x(t_i) x^3(t_j) \right ] &= \int d x(t_1)...dx(t_N)  \ x(t_i) x^3(t_j) \ p \left (x(t_1),...,x(t_i),...,x(t_j),...,...,x(t_N) \right ).
\end{align}
If these two quantities are different, the probability is not invariant under the swap of values of two arguments which implies that it is not TRS.

We test the ability of our model to learn non-TRS processes, by training it on \textit{filtered Poisson processes}. This process is defined as

\begin{align}\label{FPP_def}
\nonumber
X(t) &= \sum_k A_k  \varphi (t - t_k), \text{ where} \\
\varphi (t - t_k) &= \theta (t-t_k) e^{-(t-t_k)/\tau} \sin [\w (t-t_k)].
\end{align}
A \textit{filtered Poisson process} (FPP) $X(t)$ consists of a superposition of uncorrelated pulses $\varphi \left ( t-t_k \right )$, arriving at random times with a Poisson distriution. The overall amplitude $A_k$ is random: at each time, $A_k$ can independently take the values $\pm A$, with equal probabilities. In this process, the correlators defined in Eq.~\eqref{TRS_correlators} take the form (see the derivation in Appendix~\ref{App:FPP}) 

\begin{align}\label{correlators_exact}
\nonumber
\E \left ( X^3(t_1) X(t_2)\right ) &= \lambda I_{3,1}^{-\infty, t_1} + 3 \lambda^2 I_{1,1}^{-\infty,t_1}  I_{2,0}^{-\infty,t_1}, \\ \nonumber
\E \left ( X(t_1) X^{3}(t_2)\right ) &= \lambda I_{1,3}^{-\infty, t_1} + 3 \lambda^2 I_{1,1}^{-\infty,t_1} \left ( I_{0,2}^{-\infty,t_1} + I_{0,2}^{t_1,t_2} \right ), \\
I_{n,m}^{t, t'} &= \int_{t}^{t'} d \alpha \ \varphi^n(t_1-\alpha) \varphi^m(t_2-\alpha).
\end{align}
The two correlators are different due to the absence of TRS. We take the initial time $t =- \infty$ because we are interested in the steady state correlators.

\begin{figure}[]
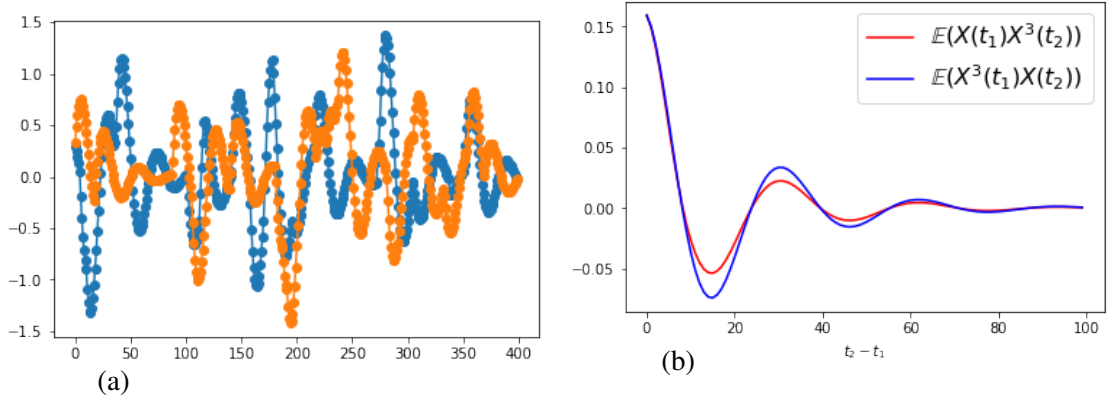

\centering
\subfigimg[width=7.4cm,angle=0]{\raisebox{-145pt}{\hspace*{35pt} (a)}}{7a}
\subfigimg[width=7.4cm,angle=0]{\raisebox{-145pt}{\hspace*{35pt} (b)}}{7b} 
\caption{(a) Training samples generated according to the FPP defined in Eq.~\eqref{FPP_def}. (b) Exact correlators of FPP defined in Eqs.~\eqref{correlators_exact} of the Poisson process defined in Eq.~\eqref{FPP_def}. The intensity of the Poisson process is $\lambda = 4$. The amplitude $A_k$ can take values $\pm 1$. The pulse decay time is $\tau = 0.2$ and the angular frequency $\w = 20$. The time indicated in the horizontal axis is an integer index that specifies the step of the discrete correlator.}
\label{FPP_exact}
\end{figure}

\begin{figure}[h]
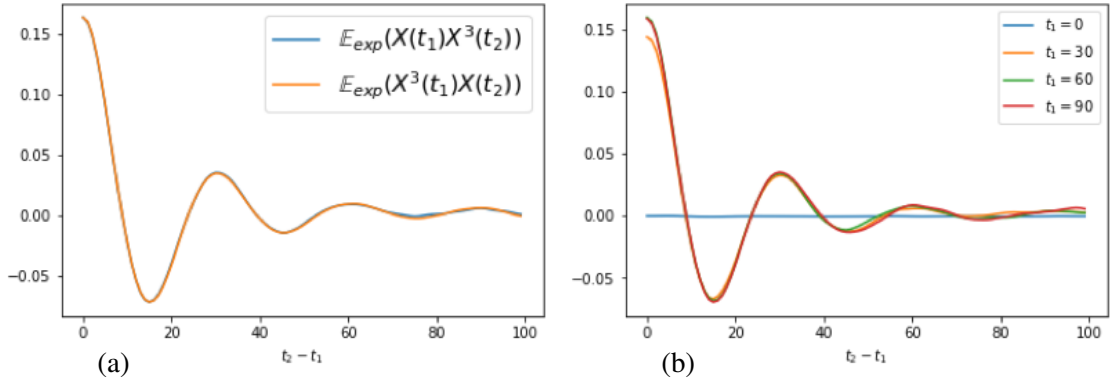

\centering
\subfigimg[width=7.4cm,angle=0]{\raisebox{-145pt}{\hspace*{35pt} (a)}}{8a}
\subfigimg[width=7.4cm,angle=0]{\raisebox{-145pt}{\hspace*{35pt} (b)}}{8b} 
\caption{The time indicated in the horizontal axis is an integer index that specifies the time step of the discrete correlator. (a) Experimental correlators $\E_{\text{exp}} \left ( X(0) X(t)^3\right )$ and $\E_{\text{exp}} \left ( X^3(0) X(t)\right )$. At $T=0.000012$, they roughly match the exact correlator $\E\left ( X^3(t_1) X(t_2)\right )$ shown in Fig.~\ref{FPP_exact}(b). The temperature can be tuned to approximately match $\E\left ( X(t_1) X^3(t_2)\right )$ instead. But at a given temperature they are both equal, unlike the exact correlators. The hyperparameters used are $dt=0.01$, $D=100$ and $\sigma = 1$. The experimental correlators are calculated by averaging over 40000 samples.
(b) The experimental correlator $\E_{\text{exp}} \left ( X(t_1) X(t_2)^3\right )$, for different values of $t_1$. It becomes stationary after $t_1 = 30$, approximately.}
\label{FPP_sample}
\end{figure}
The experimental details are shown in \ref{appendix: GP and FPP}. The dataset contains samples of the FPP defined in Eq.~\eqref{FPP_def} with parameters $A=1$ (i.e., $A_k$ can take values $\pm 1$ at time $t_k$), $\tau= 0.2$, $\w=20$. In order to create steady state signals, we produce signals of length 500, and pick the last 400 points of each signal. This corresponds to signals that have been running for a time $5 \tau$, by when signals are approximately stationary, because the process has a memory time of order $\tau$ due to the exponential decay of the pulses $\varphi(t-t_k)$. The Poisson intensity parameter is $\lambda = 4$. We show two steady state training samples in Fig.~\ref{FPP_exact}(a).

\subsubsection{Results}

By tuning the temperature, we can match the experimental correlators to either of the two exact correlators. On the other hand, both experimental correlators we obtain are equal. 

\section{Conclusions}

We introduce a quantum-inspired generative model for raw audio. It is the first machine learning model based on \textit{continuous Matrix Product States}. Our model takes the form of a stochastic Schr\"odinger equation describing the continuous time measurement of a quantum system. This constitutes a deep autoregressive architecture in which the system's state is a latent representation of past observations.

We rephrase the model in the language of stochastic differential equations. We derive an expression to calculate the two-time characteristic function of a filtered Poisson process. We test our model on three different synthetic datasets. The model is successful at learning single frequency damped sines with random delays but it fails to capture the  two frequency degree of freedom. It is able to learn \textit{Mat\'ern spectral mixtures}. Finally,  it captures the filtered Poisson process but it fails to discern between $\E \left ( X^3(t_1) X(t_2)\right )$ and $\E \left ( X(t_1) X^3(t_2)\right )$.

It remains to do a proper hyperparameter tuning considering all the hyperparameters, to see if the performance of the model can be improved. Moreover, and most importantly, the model needs to be tested on real data: how expressive is the model and how is this related to the bond dimension? How is the quantum entanglement of the model related to the the structure of correlations in the generated samples? 

This work opens a new avenue to use matrix product states to model continuous data and we hope that it will set the beginning of the exploration of cMPS for machine learning.

\acks{We are grateful to Katarzyna Macieszczak for her key insights in the design of the model. BMU and AL would like to acknowledge EPSRC under Grants EP/M506485/1 and EP/P034616/1.}

\bibliography{bibliography_MSML}

\appendix

\section{Continuous matrix product states}
\label{appendix:Continuous matrix product states}

The physical lattice states are well-captured by matrix product states. In the context of continuous quantum systems, there exist a continuum limit, without any reference to an underlying lattice parameter. This family of states are called \textit{continuous matrix product states} (cMPS). They describe field theories in one spatial dimension. In the same way MPS captures the entanglement structure of low energy states of quantum lattices, the entanglement structure of cMPS is tailored to describe the low-energy states of quantum field theories \cite{Verstraete:2010}.

For the purpose of defining a cMPS, let us consider bosons or fermions on a ring of length $L$, with field operators $\hat{\psi}(x)$ following canonical commutation relations $[\hat{\psi}(x),\hat{\psi}(y)^{\dag}]=\delta(x-y)$, with $0 \leq x,y \leq L$ space coordinates. A cMPS is defined as

\begin{equation}
\left | \Psi \right \rangle = \text{Tr}_{aux} \left [ \cP e^{\int_{0}^{L} dx \left [ Q(x) \otimes \1+R(x) \otimes \hat{\psi}(x)^{\dag}  \right ]   } \right ] \left | 0 \right \rangle.
\end{equation}
The matrices $Q(x)$ and $R(x)$ are position dependent and have dimensions $D \times D$. They act on a  $D$-dimensional ancilla. $\cP$ is the path ordering operator, $\text{Tr}_{aux}$ is the trace over the ancilla and $ \left | 0 \right \rangle$ is the vacuum state $\hat{\psi}(x) \left | 0 \right \rangle=0$. A translationally invariant state is obtained by making $Q,R$ independent of $x$. To obtain open boundary conditions, we simply substitute $\text{Tr}_{aux}$ by a left and right multiplication of the auxiliary system with a row and a column vector, respectively:

\begin{equation}
\label{cMPS_OBC}
\left | \Psi \right \rangle = \bra{v_L} \cP e^{\int_{0}^{L} dx \left [ Q(x) \otimes \1+R(x) \otimes \hat{\psi}(x)^{\dag}  \right ]   } \left | v_R \right \rangle \otimes  \left | 0 \right \rangle.
\end{equation}

\subsection{Connection between MPS and cMPS}

As shown by  Sch\"on \textit{et al}. in \cite{Schon:2005}, an MPS with bond dimension $D$ can be seen as a sequentially generated multiqubit state, arising from a $D$-level system. Let $\cH_A = \mathbb{C}^D$ and $\cH_B = \mathbb{C}^2$ be the Hilbert spaces of the ancilla and a single qubit respectively. In every step of the sequential generation, we consider unitary evolution of the joint system 
$\cH_A \otimes \cH_B$. Assuming that each qubit is initially empty $\left | 0 \right \rangle$, we disregard the qubit at the input, such that the evolution takes the form of an isometry $V: \; \cH_A \rightarrow \cH_A \otimes \cH_B$. Choosing a basis in the ancilla space, the isometry is expressed as 
\begin{equation}
\label{isometry}
V=\sum_s \sum_{a,b} A_{a,b}^s \left (\left | a \right \rangle \left \langle b \right | \otimes \left | s \right \rangle \right ), 
\end{equation}
where $\sum_s A^{s \dagger}A^{s}= \1$ is the isometry condition and each $A^s$ is a $D \times D$ matrix. After applying $V$ $n$ times to an initial state $\left |  \psi_{\text{I}} \right \rangle \in \cH_A$,
\begin{equation}
\left | \Psi \right \rangle = \sum_{\bs} \sum_{ab} A^{s_n}_{a,} ...A^{s_1}_{,b} \left \langle b  |  \psi_{\text{I}} \right \rangle
\left (\left | a \right \rangle  \otimes \left | \bs \right \rangle \right ).
\end{equation}
The generated $n$ qubits are in general entangled both with the ancilla and between themselves. If the ancilla is decoupled in the last step, the final state is an MPS in the space of the $n$ qubits:

\begin{equation}
\left | \Psi \right \rangle = \left |  \psi_{\text{F}}
 \right \rangle \otimes \sum_{\bs} \sum_{ab} \left \langle \psi_{\text{F}}  | a \right \rangle A^{s_n}_{a,} ...A^{s_1}_{,b} \left \langle b  |  \psi_{\text{I}} \right \rangle \left | \bs \right \rangle .
\end{equation}
This result shows that all sequentially generated multiqubit states, arising from a $D$-dimensional ancillary system $\cH_A$, are instances
of MPS with $D \times D$ matrices $A^s$ and open boundary conditions specified by $\left |  \psi_{\text{I}} \right\rangle$ and
$\left |  \psi_{\text{F}} \right\rangle$.

Let us now consider the cMPS shown in Eq. \eqref{cMPS_OBC}, without projecting the ancilla onto $\ket{v_L}$. Taking $L=dx$ and $Q,R$ translationally invariant, 

\begin{align}
\nonumber
\left | \Psi \right \rangle &=  \cP e^{ dx \left [ Q\otimes \1+R \otimes \hat{\psi}(x)^{\dag}  \right ]   } \left | v_R \right \rangle \otimes  \left | 0 \right \rangle \\ \nonumber
& =  \left [ \1 \otimes \1 +  Qdx \otimes \1+Rdx \otimes \hat{\psi}(x)^{\dag}  \right ]    \left | v_R \right \rangle \otimes  \left | 0 \right \rangle \\
&= \sum_{ab} \left [  \left ( \delta_{ab} + Q_{ab}dx \right ) \ket{a}\ket{b}  \otimes \1 + R_{ab} dx  \ket{a}\ket{b} \otimes \hat{\psi}(x)^{\dag} \right ]   \left | v_R \right \rangle \otimes  \left | 0 \right \rangle \\ \nonumber
&=  \sum_s \sum_{ab} A_{ab}^s \left ( \left | a \right \rangle \left \langle b \right | \otimes  \left | s \right \rangle \right ) \left | v_R \right \rangle,
\end{align}
where $A_{ab}^0 = \delta_{ab}  + Q_{ab} dx, \; A_{ab}^1 = R_{ab}dx \;{\psi^{\dagger}}^{0}(x)= \1 \; \text{and} \; {\psi^{\dagger}}^{s}(x) \ket{0} = \ket{s}$. This is just the isometry shown in Eq. (\ref{isometry}).

\section{Physical picture of cMPS}
\label{apendix:Physical picture of cMPS}

In the following we will see that a state of the form cMPS appears in the interaction picture time evolution of a composite state of a $D$-level system (which we refer to as the \textit{ancilla}), coupled to a quantum field bath. In particular we consider a $D$-level atom coupled to an electromagnetic field in the dipole approximation. The Hamiltonian of the composite system is
\begin{align}
\nonumber
H &= H_a + H_b + V, \;\; \text{where} \\ 
H_a &= \sum_n \varepsilon_n \left | n \rangle \langle n \right |, \\ 
H_b &= \sum_k \omega_k b_{k}^{\dag} b_k, \\
V &= Ep = \sum_k \left ( g_k b_k + g_k b_{k}^{\dag}\right ) \sum_{nm} p_{nm} \left | n \rangle \langle m \right |.
\end{align}
Here, $\{ \left | n \right \rangle \}$ are the $D$ eigenstates of the atom, $\{b_k\}$ are bosonic annihilation operators for each electromagnetic mode $k$ (the quantum number $k$ contains all the information specifying the mode), and $\{p_{nm}\}$ are the matrix elements of the dipole moment of the atom between different eigenstates. The coefficient $g_k$ can be assumed to be real without loss of generality and it depends on details of the electromagnetic mode $k$, specifically the volume of the space that the modes occupy \cite{Gardiner-Zoller:bookII}.

For the sake of simplicity, we will consider the case where the atom is a two-level system with energy gap $\Delta$, and so, calling the matrix element between the two levels $p_{10} \equiv p$,
\begin{equation}
V = \sum_k \left ( g_k b_k + g_k b_{k}^{\dag}\right ) \Big ( p \left | 1 \rangle \langle 0 \right | + p^* \left | 0 \rangle \langle 1 \right |  \Big ).
\end{equation}
As a first step, we go to the interaction frame with respect to $H_b$
\begin{equation}
\left | \Psi^{i}\right \rangle = U_{0}^{\dag} \left | \Psi\right \rangle, \;\; U_0 = e^{-iH_b t}.
\end{equation}
The corresponding Schr\"{o}dinger equation in the interaction picture is
\begin{equation}
\partial_t \left | \Psi^{i}\right \rangle = -i \left ( V_{IF}+H_a\right ) \left | \Psi^{i}\right \rangle,
\end{equation}
where the coupling in the interaction frame takes the form
\begin{align}
V_{IF} &= U_{0}^{\dag} V U_{0}=  \sum_k \left ( g_k b_k e^{-i \omega_k t} + g_{k} b_{k}^{\dag} e^{i \omega_k t} \right ) \left ( p \left | 1 \rangle \langle 0 \right | +
p^* \left | 0 \rangle \langle 1 \right | \right )\\
 &=  \sum_k \left ( g_k b_k e^{-i \omega_k t} + g_{k} b_{k}^{\dag} e^{i \omega_k t} \right ) \left (  e^{i \Delta t} \frac{p\left | 1 \rangle \langle 0 \right |}{e^{i \Delta t}} +
 e^{-i \Delta t} \frac{p^*\left | 0 \rangle \langle 1 \right |}{e^{-i \Delta t}} \right ).
\end{align}
Here, we introduced $1 = e^{i \Delta t}/e^{i \Delta t}$ to be able to perform a \textit{rotating wave approximation} (RWA):
\begin{equation}
V_{IF}^{RWA} = \sum_k \left (g_k  b_{k}^{\dag}  e^{-i \delta_k t} p^* \left | 0 \rangle \langle 1 \right |e^{i \Delta t}+ g_k b_k  e^{i \delta_k t} p \left | 1 \rangle \langle 0 \right |e^{-i \Delta t}\right ),
\end{equation}
where $\delta_k \equiv \Delta-\omega_k$ is the detuning. We define the bath operator $b(t)\equiv e^{-i \Delta t} \sum_k g_k b_k e^{i \delta_k t}$ and raising operator $R^{\dag} \equiv i p \left | 1 \rangle \langle 0 \right |$.  The resulting Schr\"{o}dinger equation takes the form
\begin{equation}
\partial_t \left | \Psi^{i}\right \rangle = \left ( Rb^{\dag}(t)-R^{\dag}b(t)-iH_a\right ) \left | \Psi^{i}\right \rangle .
\end{equation}
The time dependence of $b(t)$ stems not only from the fact that we are in the interaction frame but also the $e^{i \Delta t}$ we introduced to perform the RWA. This is equivalently an atomic system in the Schr\"{o}dinger picture, where the system is driven by these fields, which are regarded as known time-dependent operators \cite{Gardiner-Zoller:bookII}. 

These new operators do not follow bosonic commutation relations, instead
\begin{equation}
[b(t),b^{\dag}(t')] = e^{-i \Delta (t-t')} \sum_k g_{k}^{2}  e^{i \delta_k (t-t')}.
\end{equation}
For certain baths, $ \sum_k g_{k}^{2}  e^{i \delta_k (t-t')}$ is sharply peaked at $t=t'$ \cite{Wiseman-Milburn}. Therefore, we will approximate this function with a delta function:
\begin{equation}
[b(t),b^{\dag}(t')] = e^{-i \Delta (t-t')}  \delta(t-t') =  \delta(t-t').
\end{equation}
This corresponds to taking the \textit{Markovian} limit \cite{Wiseman-Milburn}. In the remainder of the derivation, we define the differential bath operator $dB_t\equiv b(t)dt$. We note that
\begin{equation}
\label{dB_comm}
[dB_t,dB^{\dag}_{t}]=\underbrace{\delta(0)dt}_{1}dt=dt.
\end{equation}
and so  $dB_t\sim \sqrt{dt}$. This can be understood by thinking of $dt$ as the smallest unit into which time can be devided. Then we have a discrete delta function with finite width and height, with area $\delta(0)dt=1$.

We now consider the case where the electromagnetic field is in the vacuum state $\ket{0}$. Considering a differential step and expanding to order $dt$
\begin{align}
\nonumber
\left | \Psi^{i}_{dt}\right \rangle &= \exp \left ( R dB^{\dag}_{dt}- R^{\dag} dB_{dt}-iH_a dt \right )  \left | \psi^{i}_{0} \right \rangle \otimes \left | 0 \right \rangle \\
&\approx \left ( \1- \left ( iH_a +\frac{R^{\dag}R}{2} \right  )dt +RdB^{\dag}_{dt} + \frac{R^2}{2} dB^{\dag}_{dt}dB^{\dag}_{dt}\right ) \left | \psi^{i}_{0} \right \rangle \otimes \left | 0 \right \rangle.
\end{align}

Note that $dB_{dt} \left |0 \right \rangle=0$ and $dB_{dt}dB^{\dag}_{dt}\left |0 \right \rangle=dt \left |0 \right \rangle$ from Eq.~\eqref{dB_comm}. Neglecting the last term, this is the first order expansion of the continuous matrix product state defined in Eq.~\eqref{cMPS_OBC}, given we identify $Q = -iH_a - R^{\dagger}R/2$ and $\hat{\psi}^{\dagger} dt = dB^{\dagger}$. A more careful analysis (beyond the scope of this work) reveals that the last term $dB^{\dagger} dB^{\dagger}$ need not be kept \cite{Gardiner-Zoller:bookII}. Thus we are finally left with a \textit{continuous matrix product state} \cite{Osborne:2010, Verstraete:2010}
\begin{align}
\left | \Psi^{i}_{dt}\right \rangle &\approx \left [ \1- \left ( iH_a+\frac{R^{\dag}R}{2} \right  )dt +RdB^{\dag}_{dt} \right ] \left | \psi^{i}_{0} \right \rangle \otimes \left | 0 \right \rangle.
\end{align}
As a last step let us consider the time evolution of $\left | \Psi^{i'}_{dt}\right \rangle = e^{i H_a t}\left | \Psi^{i}_{dt}\right \rangle$ so that the model takes a more compact form. Then,

\begin{align}
\label{cMPS}
\left | \Psi^{i'}_{t+dt}\right \rangle &= \left [ \1- \frac{R_{t}^{\dag}R_{t}}{2} dt +R_{t}dB^{\dag}_{t+dt} \right ] \left | \psi^{i'}_{t} \right \rangle \otimes \left | 0 \right \rangle,
\end{align}
where $R_{t}=e^{i H_a t} R e^{-iH_a t}$. In the remainder, we will not keep the $i'$ index but we will still be referring to states whose time evolution is \eqref{cMPS}. Also we will use $H$ to refer to $H_a$.

\subsection{Balanced homodyne detection}
\label{appendix:Balanced homodyne detection}

Balanced homodyne measurement corresponds to mixing of the output field with a strong (classical) oscillator (mode $a$) on a balanced beam splitter, and measuring the photon number difference between the two output fields $c=\ (a+b)/\sqrt{2} $ and $d=(a-b)/\sqrt{2}$:

\begin{equation}
\Delta n = c^\dagger c-d^\dagger d= a b^\dagger + a^\dagger b\approx \alpha b^\dagger + \alpha^* b,
\end{equation}
where the last approximation follows from the operator $\Delta n$ acting on the coherent state $|\alpha\rangle$, i.e., $a|\alpha\rangle\approx\alpha |\alpha\rangle$. In particular, the approximation becomes exact for the photon count divided by the oscillator amplitude in the strong oscillator limit,

\begin{equation}
I \equiv \lim_{|\alpha|\rightarrow\infty}\frac{c^\dagger c-d^\dagger d}{|\alpha|}= e^{i\phi} b^\dagger + e^{-i\phi}  b. \label{hom0}
\end{equation}

We now discuss the effect of the operator $I = e^{i\phi} b^\dagger + e^{-i\phi}  b$ being measured continuously on the output of an open quantum system described by a cMPS. As shown in Eq. \eqref{cMPS},

\begin{eqnarray}
\label{hom:Psi}
|\Psi_{t+dt}\rangle &=&\left[\mathds{1}- \frac{1}{2} R_{t}^\dagger R_t dt+ R_t \otimes dB^{\dagger}_{t+dt}\right] |\psi_t \rangle\otimes|0\rangle\\
&=& \left[\mathds{1}- \frac{1}{2} R_{t}^\dagger R_t dt+ R_t \otimes (dB^{\dagger}_{t+dt}+ e^{-i2\phi}dB_{t+dt}) \right] |\psi_t\rangle\otimes|0\rangle, \nonumber
\end{eqnarray}
where we introduced $e^{-i2 \phi}dB_{t+dt}$ so that we can introduce the operator $I$ in the equation of motion. If we make a measurement of $I$ at time $t+dt$, projecting the state $|\Psi_{t+dt}\rangle$ onto $\ket{I_{t+dt}} \otimes \langle I_{t+dt}|\Psi_{t+dt}\rangle$, we are left with the following state of the ancilla

 \begin{equation}
 \langle I_{t+dt}|\Psi_{t+dt}\rangle \equiv  |\widetilde\psi_{t+dt}\rangle = \left[\mathds{1}- \frac{1}{2} R_{t}^\dagger R_{t}  dt+ R_{t}\, e^{-i\phi} I_{t+dt} dt \right] | \psi_t\rangle \times \sqrt{\mathcal{P}(I_{t+dt})},\label{hom:psitilde}
 \end{equation}
where  $\mathcal{P}(I_{t+dt})= |\braket{0|I_{t + dt}}|^2 =\sqrt{dt/2\pi}\exp(- dt I^{2}_{t+dt} /2) $ is the probability of measuring $I_{t+dt}$ on the vacuum state and

\begin{eqnarray}
p(I_{t+dt})  =\langle \widetilde\psi_{t+dt} |\widetilde\psi_{t+dt}\rangle& =&  \mathcal{P}(I_{t+dt})  \, \bigg \{1 +  \langle e^{-i\phi} R_{t} + e^{i\phi} R_{t}^\dagger \rangle_{\psi_t} I_{t+dt} dt\label{hom:psinorm} \\
&&+\left[- 1 + \left(I_{t+dt} \sqrt{dt}\right)^2 \right]\langle R_{t}^\dagger R_{t} \rangle_{\psi_t} \, dt +\mathcal{O}(I_{t+dt} dt^2, dt^2)   \bigg\}\nonumber
\end{eqnarray}
is the probability density of obtaining $I_{t+dt}$. Recalling Eqs.~\ref{dB_comm} and \ref{hom0}, note that $I \sim 1/\sqrt{dt}$. Then $\left[- 1 + \left(I_{t+dt} \sqrt{dt}\right)^2 \right]$ is of order one and

\begin{equation}
p(I_{t+dt}) =  \sqrt{\frac{dt}{2\pi}}\exp\left[- \frac{dt}{2} I_{t+dt}^2 + \langle e^{-i\phi} R_{t} + e^{i\phi} R_{t}^\dagger \rangle_{\psi_t} I_{t+dt} dt +\mathcal{O}(dt)\right].  
\end{equation}
We now add a term of order $dt$ to complete the square so that we get a Gaussian probability density

\begin{align}
\nonumber
p(I_{t+dt}) &=  \sqrt{\frac{dt}{2\pi}}\exp\left[- \frac{dt}{2} {\left (I_{t+dt} - \langle e^{-i\phi} R_{t} + e^{i\phi} R_{t}^\dagger \rangle_{\psi_t} \right )^2} +\mathcal{O}(dt)\right] \\  \nonumber
& \approx  \sqrt{\frac{dt}{2\pi}}\exp\left[- \frac{dt}{2} {\left (I_{t+dt} - \langle e^{-i\phi} R_{t} + e^{i\phi} R_{t}^\dagger \rangle_{\psi_t} \right )^2} \right] \\ 
&=   \sqrt{\frac{1}{2\pi \left( 1/ \sqrt{dt}\right )^2}}\exp\left[- \frac{\left (I_{t+dt} - \langle e^{-i\phi} R_{t} + e^{i\phi} R_{t}^\dagger \rangle_{\psi_t} \right )^2}{2 \left( 1/ \sqrt{dt}\right )^2} \right].
\end{align}
Equivalently,

\begin{equation}
I_{t+dt} =  \langle e^{-i\phi} R_{t} + e^{i\phi} R_{t}^\dagger \rangle_{\psi_t}+ z, \text{ where } z  \sim N(0,1/dt),
\end{equation}
where $N(0,1/dt)$ is a Gaussian distribution with zero mean variance $1/dt$. For the remainder of the chapter we fix $\phi=0$. The conditional joint probability density for a sequence of measurements $\{ I_t\}$ is

\begin{align}
\nonumber
p(I_T, ..., I_1 |H,R) &= \prod_{k=0}^{T-1} p(I_{k+1}| I_k, ...,I_1; H,R), \text{ where} \\
p(I_{k+1}| I_k, ...,I_1; H,R) &=   \sqrt{\frac{1}{2\pi \left( 1/ \sqrt{dt}\right )^2}}\exp\left[- \frac{\left (I_{k+1} - \langle R_{t} + R_{t}^\dagger \rangle_{\psi_k} \right )^2}{2 \left( 1/ \sqrt{dt}\right )^2} \right],
\end{align}
where $\psi_k$ is the state of the ancilla at time $k$.

Aside from a few last details and refinements that are described in the main text, this probability distribution defines our \textit{quantum-inspired model}.

\section{Experimental details}
\label{appendix: experimental details}
\subsection{Damped sines}
\label{appendix: damped sines}

\begin{enumerate}

	\item We consider the time derivative of the data as the homodyne current, i.e., $I_t = dx_t /dt$. The matrix $R$ is complex and we set its diagonal elements to zero. Hence, we only keep oscillatory parts of $R(t)$, which we consider appropriate to model oscillatory data.
	
	\item We learn the initial state $\ket{\psi_0}$ (or $\rho_0$). When using density matrices, we parameterize $\rho_0$ by $\rho_0 = \frac{W^\dagger W}{\tr\left[W^\dagger W\right]}$ to enforce normalization and real and positive eigenvalues. The matrix $W\in\mathbb{C}^{r\times D}$ defines the rank of the initial density matrix, with $r=1$ corresponding to an initial pure state $\ket{\psi_0}$.
	
	\item We use regularisers for the elements of $H$ and $R$. These are set to $\sigma_{\w}^2 = \frac{(16000 \pi)^2}{400}$ and $\sigma_R^2 = 5$, defined in Appendix \ref{appendix:regularization}. They are included in the model as a refinement, but we do not experiment with them. They will become important when training on real data, which is more complex than the data considered here. 
	
	\item The hyperparameter $dt$ remains fixed to $dt = 1/16000$. We experiment with different values of $D$ and $\sigma$ but only show results with the values that give the best results.
	
	\item The batch size is 8.
\end{enumerate}

\subsection{GP and FPP}
\label{appendix: GP and FPP}

\begin{enumerate}
\item We consider the data to be the homodyne current, i.e. $I_t = x_t$. This is because as shown in Fig.~\ref{x_vs_dx}, on this dataset the increments of the signal are a lot more spiky than the signals themselves, which makes learning difficult in the continuous formulation of the problem.
\item We learn the initial state $\ket{\psi_0}$.
\item The hyperparameter $dt$ is set to $dt=0.001$ and $\sigma=1$. We experiment with different values of $D$ and $\sigma$ but only show results with the values that give the best results.
\item We do not use regularisers.
\item The batch size is 8.
\end{enumerate}

\section{Details about Gaussian Processes}
\label{appendix: details about GP}

A stochastic function $x(t)$ is a Gaussian process (GP) if any finite collection of random variables $x(t_1),...,x(t_n)$ have a multidimensional Gaussian distribution \cite{Sarkka:2019}. A GP is defined in terms of its \textit{mean} m$(t)$ and its \textit{covariance function} (or \textit{kernel}) C$(t,t')$, defined as

\begin{align}
\text{m}(t) &=  \E [x(t)], \\
\tC(t,t') &=  \E \left [ \left ( x(t)-\text{m}(t) \right ) \left ( x(t')-\text{m}(t') \right ) \right ].
\end{align}
A Gaussian process is \textit{stationary} if the mean is time independent and the covariance function only depends on time differences

\begin{equation}
\tC(t,t') = \tC(t-t').
\end{equation}
We use the notation $\tC(\tau)$ (where $\tau \equiv t-t'$) when considering stationary processes. The Wiener-Khintchine theorem relates the stationary kernel to a corresponding spectral function

\begin{align}
S(\w) &= \int_{-\infty}^{\infty} d \tau \ \tC(\tau) e^{-i \w \tau}, \\
\tC(\tau) &= \frac{1}{2 \pi} \int_{-\infty}^{\infty} d \w \ S(\w) e^{i \w \tau}.
\end{align}
A specific kind of stationary GPs that have been used to reflect the complex harmonic structure of musical notes are \textit{Mat\'ern spectral mixtures}. They have been used for different sound related machine learning tasks \cite{Alvarado:2017}. Consider the kernels

\begin{align}
\tC_{1/2}(\tau) &= \sigma^2 e^{- \lambda \tau} ,\\
\tC_{\cos}(\tau) &= \cos (\w_0 \tau).
\end{align}
The corresponding spectral densities are

\begin{align}
S_{1/2}(\w) &= \frac{2 \sigma^2 \lambda}{\lambda^2 + \w^2}, \\
S_{\cos}(\w) &= \pi \left [  \delta (\w - \w_0) + \delta(\w + \w_0) \right ].
\end{align}
The spectral density of the product of the two kernels $C(\tau) = C_{1/2}(\tau)C_{\cos}(\tau)$ takes the form of a pair of Lorentzians centered at $\pm \w_0$

\begin{equation}
S(\w; \boldsymbol{\theta}) = 2 \pi \sigma^2 \lambda \left [   \frac{1}{\lambda^2 + (\w-\w_0)^2} +  \frac{1}{\lambda^2 + (\w+\w_0)^2} \right ],
\end{equation}
where $\boldsymbol{\theta} = (\sigma, \lambda, \w_0)$. The general form of \textit{Mat\'ern spectral mixtures} is a sum over different pairs of Lorentzians 

\begin{equation}
S_{\text{SMS}} (\w ; \boldsymbol{\Theta}) = \sum_{j=1}^{N} S(\w; \boldsymbol{\theta}_j),
\end{equation}
where $\boldsymbol{\Theta} = \{\boldsymbol{\theta}_j \}$. The corresponding covariance is

\begin{equation}\label{MSM}
\tC_{\text{MSM}} (\tau; \boldsymbol{\Theta}) = \sum_{j=1}^N \sigma_j^2 e^{- \lambda_j \tau} \cos(\w_j \tau).
\end{equation}

\section{Two frequencies experiment}
\label{appendix: Two frequencies experiment}

We want to see if we can generate samples of two different frequencies, after training on a dataset of damped sines with random delays and two different frequencies $f=600,800$Hz. The length of the training sequences is 100 samples.

\begin{figure}[h]
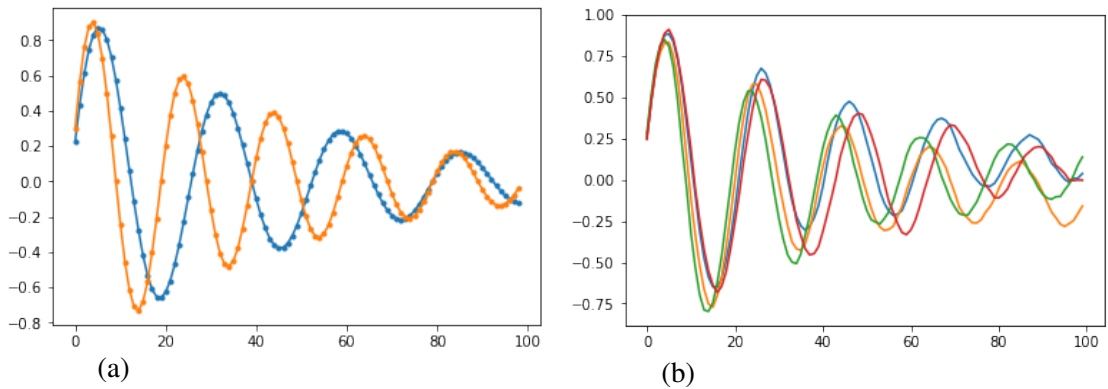

\centering
\subfigimg[width=7.4cm,angle=0]{\raisebox{-145pt}{\hspace*{35pt} (a)}}{9a} 
\subfigimg[width=7.4cm,angle=0]{\raisebox{-145pt}{\hspace*{35pt} (b)}}{9b}
\caption{(a) Training set made of two signals. (b) Four samples at T=100 and $D=50$ and $\sigma = 10^{-4}$, after having trained on the two signal dataset shown in (a).}
\label{2_freq_fix}
\end{figure}

We start with the pure state model. We train the model on a dataset that only contains two signals, shown in Fig.~\ref{2_freq_fix}(a). 
After training, our model generates signals with different frequencies that lie in between the two frequencies of the dataset, as shown in Fig.~\ref{2_freq_fix}(b). The frequencies of the samples seem to be closer to $f=800$Hz than to $f=600$Hz.

\begin{figure}[h]
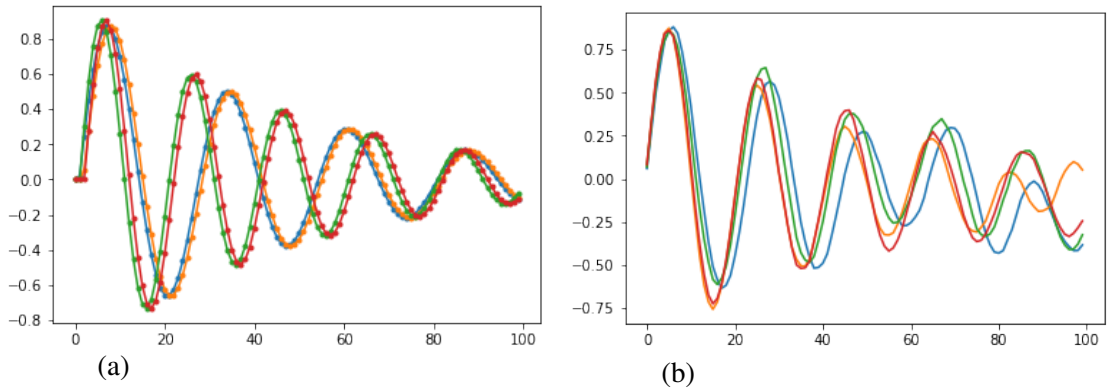

\centering
\subfigimg[width=7.4cm,angle=0]{\raisebox{-145pt}{\hspace*{35pt} (a)}}{10a} 
\subfigimg[width=7.4cm,angle=0]{\raisebox{-145pt}{\hspace*{35pt} (b)}}{10b}
\caption{(a) Four signals of the dataset with different delays and frequencies $f=600,800$ Hz. The length of the data samples is 100 and the sampling frequency 16 kHz. (b) Four samples at $T=120, D=100$ and $\sigma = 10^{-4}$.}
\label{2freq_delay}
\end{figure}

We then move on to modeling a dataset of damped sines with random delays like we did in Sec.~\ref{1_freq_exp}, but this time the training set will contain damped sines of two different frequencies as shown in Fig.\ref{2freq_delay}(a). We find that after training, samples are always quite close to the higher frequency. Different generated samples have different shapes, but all look like damped sines. The model learns the manifold of damped sines fairly well, but it fails to capture the two frequencies degree of freedom of the dataset, in that there are no samples with frequencies close to $f=600$Hz. We show the result in Fig.~\ref{2freq_delay}(b).

We experiment considering the time evolution of a density matrix but the performance of the model does not improve compared to the pure state case, i.e.  it fails to capture the two frequencies degree of freedom of the dataset.

\section{Regularization}
\label{appendix:regularization}

What should be the range of values of the learnt parameters? If we had any intuition or knowledge about this question, we could use it to bias the learning. The way to bias or constrain learning is to introduce regularizers. This is equivalent to introducing a prior and doing maximum a posteriori instead of maximum likelihood.

\textbf{Regularization of $H$}

Consider the following discretized Scr\"odinger equation in the interaction picture

\begin{equation}
\label{SSE_Schr}
\ket{\tilde\psi_{t+\Delta t}} = \left[\1-\frac{\sigma^2}{2}R^\dagger(t) R(t)\Delta t+R(t) \Delta x_t\right]\ket{\tilde \psi_t}
\end{equation}
where $R(t) \equiv e^{iHt} R e^{-iHt}$. If $H$ is diagonal with eigenvalues $\omega_n$, the matrix elements of $R$ are $R_{ab}(t)=R_{ab} e^{i(\omega_a - \omega_b)t}$. Suppose we want to learn a single sequence, e.g., $x_t = \sin(\w t)$ [i.e., $ \Delta x_t / \Delta t \approx \w \cos(\w t)$], such that the loss function is

\begin{equation}
\text{loss} =  \sum_t 
\left (\omega \cos(\w t) - A \langle R(t)+R^\dagger(t)\rangle_t  \right )^2.
\end{equation}
The loss function is minimised when $A \langle R(t)+R^\dagger (t)\rangle_t = \omega\cos(\omega t)$, i.e. when the expectation value $ \langle R(t)+R^\dagger (t)\rangle_t$ oscillates with frequency $\w$. Since matrix elements of both $R(t)$ and $R^\dagger(t) R(t)$ oscillate with frequencies that are differences of eigenvalues of $H$, it is intuitive that the learned diagonal elements of $H$ should be related to the frequency $\w$ of the training data. 

If we assume that $H$ is related to the frequencies, it makes sense to limit it to the bandwidth of audio. \textit{Nyquist's theorem} states that in order to correctly capture a discrete signal, the sampling rate must be at least double the highest frequency contained in the signal. Conversely, the highest frequency that can be captured at a given sampling rate is half the sampling frequency. This frequency is called the \textit{Nyquist frequency}. If differences of eigenvalues of $H$ give frequencies, the spectrum of $H$ should be limited to $ \pm s/4$, where $s$ is the sampling rate. Thus if we set the standard deviation of the frequencies to be $\sigma_{f}=s/4$, and bearing in mind that $\omega = 2\pi f$, the regularization term in the loss should be

\begin{equation}
\mathcal{L}_H = \frac{1}{2\sigma_\omega^2}\sum_n \w_n^2= \frac{1}{8\pi^2\sigma_f^2}\sum_n \w_n^2=\frac{2}{\pi^2s^2}\sum_n \w_n^2,
\end{equation}
which, up to a constant, corresponds to the logarithm of the Gaussian prior

\begin{equation}
p(\w_1, ..., \w_D) = \prod_{n=1}^{D} \sqrt{\frac{1}{2 \pi \sigma_{\w}^{2}}} \exp \left ( -\frac{\w_n^2}{2 \sigma_{\w}^{2}}  \right ).
\end{equation}

\textbf{Regularization of $R$}

The scale of the signal  is set by $A \langle R+R^\dagger \rangle_t$. If the typical scale of the matrix elements of $R$ is $r$, its value should be determined by $\Delta x = A r \Delta t$. If we set $A=1$, we could introduce a Gaussian prior so that $\sigma_R = \Delta x / \Delta t$. The hyperparameter $\Delta x$ can be inferred from the data. Then,

\begin{equation}
\mathcal{L}_R = \frac{1}{2\sigma_R^2}\sum_{ij} |r_{ij}|^2. 
\end{equation}
On the other hand, in general $A$ is a learning variable and so it is not obvious what the regularizer of $R$ should be.

\section{Relation between covariance functions and SDEs}
\label{Relation between covariance functions and SDEs}

Most of the theory explained in this appendix can be found in Chapters 6 and 12 of \cite{Sarkka:2019}. A stochastic differential equation (SDE) is a differential equation that contains terms which are random functions. This implies that their solutions are also random functions. Consider a Gaussian noise-driven ordinary differential equation of the form

\begin{equation}\label{SDE}
d \bx = \textbf{f} (\bx,t) dt + \bL(\bx, t) d\boldsymbol{\beta}(t),
\end{equation}
where $\boldsymbol{\beta}(t)$ is Brownian motion with diffusion matrix $\bQ$ and $ \textbf{f} (\bx,t)$ and $\bL(\bx, t)$ are arbitrary vector- and matrix-valued functions, respectively. The solutions $\bx(t)$ of SDEs are random processes and therefore they have certain probability distribution $p(\bx(t))$ [also denoted $p(\bx, t)$]. This probability density solves the \textit{Fokker--Planck--Kolmogorov} (FPK) equation

\begin{align}\label{FPK}
 \nonumber
\frac{\partial p(\bx,t)}{\partial t} =& - \sum_i \frac{\partial}{\partial x_i} \left [ f_i (\bx, t) p(\bx, t) \right ] \\
&+ \frac{1}{2} \sum_{i,j} \frac{\partial^2}{\partial x_i \partial x_j} \left \{ \left [ \bL (\bx,t) \bQ \bL^T(\bx,t)\right ]_{ij}  p(\bx,t) \right \},
\end{align}
given the initial condition $p(\bx, t_0)$. One can obtain the equations of motion for the mean, covariance and other statistical quantities from this equation. Among others, denoting the mean \textbf{m}$(t)= \E [\bx(t)]$ and the \textit{marginal covariance} \textbf{C}$(t,t)\equiv$\textbf{P}$(t)=  \E \left [ \left ( \bx(t)-\tbm(t) \right ) \left ( \bx(t)-\tbm(t) \right )^T \right ]$, 

\begin{align}
\frac{d \textbf{m}}{dt} &= \E \left [ \bbf (\bx, t)\right ], \\
\frac{d \textbf{P}}{dt} &= \E \left [ \bbf (\bx, t) (\bx -\textbf{m})^T \right ] + \E \left [ (\bx -\textbf{m})  \bbf^T (\bx, t) \right ]
+\E \left [ \bL (\bx,t) \bQ \bL^T(\bx,t)\right ] .
\end{align}
Another useful quantity that we can obtain from the FPK equation~\eqref{FPK} is the \textit{transition density} $p(\bx(t)|\bx(s))$ of the SDE in \eqref{SDE}, which is the probability of the random process taking the value $\bx(t)$ at time $t$, given the value at time $s$ was $\bx(s)$. This quantity is the solution of the FPK equation \eqref{FPK}, with the initial condition $p(\bx(t)|\bx(s))=\delta (\bx (t)-\bx(s))$ at $t=s$. 

An SDE is \textit{linear} if $\bbf = \bF \bx$. The covariance function \textbf{C}$(t,t')$ of linear stochastic differential equations can be obtained from the marginal covariance

\begin{equation}
\textbf{C}(t,t') =
\begin{cases}
\bP(t) \exp \left [ (t'-t) \bF \right ]^T, & \text{if }t<t', \\
\exp \left [ (t-t') \bF \right ] \bP(t'), & \text{if } t\geq t'.
\end{cases}
\end{equation}

\subsection{Equivalent discretisations of linear time-invariant SDEs}
\label{equivalent_disc}

An SDE is \textit{time-invariant} if $\bbf$ and $\bL$ do not depend on time. Consider the linear time-invariant stochastic differential equation

\begin{equation}
d \bx = \textbf{F}  \bx dt + \bL d\boldsymbol{\beta},
\end{equation}
with initial conditions $\bx(t_0)\sim \text{N}(\textbf{m}_0, \textbf{P}_0)$, where $\text{N}(\textbf{m}_0, \textbf{P}_0)$ denotes a Gaussian distribution with mean $\textbf{m}_0$ and marginal covariance $\textbf{P}_0$. From the FPK equation, one obtains the transition density

\begin{equation}\label{trans}
p \left (\bx(t)| \bx (s) \right )=N \left (\mathbf{m}(t|s),\bP(t|s) \right ),
\end{equation}
where

\begin{align}
\mathbf{m}(t|s)&=\exp \left ( \mathbf{F}(t-s) \right ) \bx(s), \\
\bP(t|s) &= \int_{s}^{t} \exp \left ( \mathbf{F}(t-\tau) \right ) \bL \bQ \bL^T \exp \left ( \mathbf{F}(t-\tau) \right )^T d \tau.
\end{align}
Let us consider discrete times $\{ t_k\}$, separated by $\Delta t_k$.  Eq.~\eqref{trans} then implies

\begin{equation}
\bx(t_{k+1})-\mathbf{m}(t_{k+1}|t_k) = \bq_k, \;\;\; \bq_k \sim N(0,\bP(t_{k+1}|t_k)).
\end{equation}
Therefore, we derive a discrete stochastic equation

\begin{equation}\label{SDE_discrete}
\bx(t_{k+1})= \exp \left ( \mathbf{F}(\Delta t_k) \right ) \bx(t_k) + \bq_k, \;\;\; \bq_k \sim N(0,\bP(\Delta t_k|0)).
\end{equation}
This discretization is exact in that the probability distribution of the continuous and discrete models defined in Eqs.~\eqref{SDE} and \eqref{SDE_discrete}, coincide at times $\{ t_k\}$.

\subsection{From steady state covariance functions to discrete stochastic processes}

As shown at the beginning of Appendix~\ref{Relation between covariance functions and SDEs}, it is possible to derive the covariance function of an SDE. Conversely, it is also possible to find the SDE that corresponds to a given covariance function. Consider the following steady state covariance

\begin{align}
&\bC(\tau) = \bC_{\cos}(\tau) \bC_{\exp}(\tau), \; \text{where} \\ \nonumber
&\bC_{\cos}(\tau)=\cos(\omega \tau), \\ \nonumber
&\bC_{\exp}(\tau)=\sigma^2  e^{-\lambda |\tau|}. 
\end{align}
As shown in \cite{William:2018, Solin:2014}, the corresponding SDE is

\begin{align}\label{continuous_SSM} 
\nonumber
d \bg(t) &= \textbf{F}  \bg(t) dt + \bL d\boldsymbol{\beta}, \\
x(t) &= \bH \bg (t),
\end{align}
where

\begin{align}
\bF&=
\begin{pmatrix}
-\lambda & -\w \\
\w & -\lambda
\end{pmatrix}, \\
\bL &=  \1_2, \\
\bH &= (1,0).
\end{align}

This is called a \textit{continuous state space model}, the vector $\bH$ is the \textit{measurement model} and $\bg(t)$ the \textit{state}. The equivalent discretization of Eq.~\eqref{continuous_SSM} is

\begin{align}\label{discrete_SSM}
\mathbf{g}_{k+1} &= \mathbf{A} \mathbf{g}_k + \bq_k, \;\;\; \bq_k \sim N(0,\bm\Sigma), \\
x(t_k) &= \bH \bg_k,
\end{align}
where

\begin{align}
\bA&=\exp(-\lambda \Delta t)
\begin{pmatrix}
\cos(\omega \Delta t) & -\sin(\omega \Delta t) \\
\sin(\omega \Delta t) & \cos(\omega \Delta t)
\end{pmatrix}, \\
\bm\Sigma &= \sigma^2 (1-e^{-2 \lambda \Delta t}) \1_2.
\end{align}
When the kernel is a sum of $N$ stationary kernels $\bC(\tau) =\sum_{i=1}^{N} \bC_i (\tau)$, we get the corresponding SDE by replacing $\bF,\bL$ and $\bH$ in Eq.~\eqref{continuous_SSM} by

\begin{align}
\bF &= \text{blkdiag}(\bF_1,...,\bF_N), \\
\bL &= \text{blkdiag}(\bL_1,...,\bL_N), \\
\bQ &= \text{blkdiag}(\bQ_1,...,\bQ_N), \\
\bH &= (\bH_1,...,\bH_N ). 
\end{align}
Here $\bQ$ is the diffusion matrix of Brownian motion $\bm\beta(t)$. The corresponding equivalent discretization is obtained by performing the equivalent substitution of $\bA,\bm\Sigma$ and $\bH$ in Eq.~\eqref{discrete_SSM}. The dimension of the state vector $\bg$ in Eqs.~\eqref{continuous_SSM} and \eqref{discrete_SSM} is then increased by a factor of $N$.

Obtaining samples from these discrete stochastic processes is equivalent to sampling the corresponding multidimensional Gaussian distributions.

\section{Filtered Poisson processes}
\label{App:FPP}

\subsection{Poisson process}

A \textit{standard Poisson process} N$_t$ is a counting process that has jumps of size $+1$ at homogeneously distributed random times and its path is constant in between two jumps. This is defined as

\begin{equation}
N_t = \sum_{k=1}^{\infty} \1_{[t_k, \infty)}, \;\; \text{for } t \geq 0,
\end{equation}
where

\begin{equation}
 \1_{[t_k, \infty)} =
\begin{cases}
1, & \text{if }t \geq t_k, \\
0, & \text{if } 0 \leq t < t_k.
\end{cases}
\end{equation}

Furthermore, a Poisson process satisfies the following conditions:

\begin{enumerate}
  \item Independence of increments: for all $0 \leq t_0 < t_1 < \cdot \cdot \cdot < t_n$, the increments
  
\begin{equation}
N_{t_1}-N_{t_0}, ..., N_{t_n}-N_{t_n -1},
\end{equation}
are independent random variables.

\item Stationarity of increments: $N_{t+h}-N_{s+h}$ and $N_{t}-N_{s}$ have the same distribution for all $h>0$ and $0 \leq s \leq t$. 

\item Conditions 1 and 2 imply that the probability distribution of the increments is a Poisson distribution, i.e. for all $0 \leq s \leq t$, 

\begin{equation}
p(N_t - N_s = k) = e^{- \lambda (t-s)} \frac{(\lambda (t-s))^k}{k!}.
\end{equation}
The parameter $\lambda$ is called the intensity of the Poisson process.

\end{enumerate}
From the last condition we can infer the sort time asymptotics

\begin{align}
\nonumber
p(N_{\Delta t} =0) &= e^{-\Delta t \lambda} = 1 - \Delta t \lambda + \cO (\Delta t^2) \approx 1 - \Delta t \lambda ,  \\
p(N_{\Delta t} =1) &= \Delta t \lambda e^{-\Delta t \lambda} = \Delta t \lambda + \cO (\Delta t^2) \approx \Delta t \lambda, \;\; \Delta t \rightarrow 0.
\end{align}

\subsection{One-time characteristic function of a filtered Poisson process}

A \textit{filtered Poisson process} (FPP) $X(t)$ consists of the superposition of uncorrelated pulses $\varphi \left ( t-t_k \right )$, where the arrival times $\{ t_k \}$ follow a Poisson distribution

\begin{equation}\label{FPP}
X(t)=\sum_{k} A_k \varphi \left (t-t_k \right ).
\end{equation}
The overall amplitude $A_k$ is random. Let us consider the case where at each time, $A_k$ can independently take the values $\pm A$, with equal probabilities. The characteristic function of $X(t)$ is

\begin{equation}\label{ch_func}
\Phi_X (u,t) = \E \left (  e^{i u X(t)} \right ),
\end{equation}
where the average is taken over all possible Poisson processes. Note that this involves averaging over the random set of jump times $\{ t_k\}$ as well as the value of the sequence of amplitudes $\{ A_k \}$ at times $\{ t_k\}$. Using Campbell's theorem \cite{wiki:Campbells_theorem},

\begin{align}\label{Campbell}
 \E \left (  e^{i u X(t)} \right ) &= \exp \left \{ - \lambda \int_{0}^{t} d \alpha \left [ 1 - \Phi_A (i u \varphi (t- \alpha)) \right ] \right \},
 \text{ where} \\
  \Phi_A (i u \varphi (t- \alpha)) &=  \E_A \left (  e^{i u A \varphi (t- \alpha)} \right ).
\end{align}

\underline{\textit{Naive proof}}:

Let us rewrite the process as a sum over all time steps $\{ i \Delta t \}$, instead of the jump times $\{ t_k \}$ only

\begin{equation}
X(t)=\sum_{k \in \{t_k \}} A_k \varphi \left (t-t_k \right ) = \sum_{i=1}^{N} \sigma_i A_i \varphi \left (t-i \Delta t \right ),
\end{equation}
where $t=N \Delta t$. The parameter $\sigma_i$ is 1 if there is a jump, and 0 otherwise. The expectation value in \eqref{ch_func} is an average over the random variables $A$ and $\sigma$ at each time step, i.e.,

\begin{equation}
\Phi_X (u,t) = \E \left (  e^{i u X(t)} \right ) = \E_{\sigma} \left [ \E_{A} \left (  e^{i u X(t)} \right ) \right ].
\end{equation}
Let us define

\begin{equation}
\Phi_{A_i}\left [u \varphi (t- i \Delta t) \right ] \equiv \E_{A_i} \left (  e^{i u \sigma_i A_i \varphi (t- i \Delta t)} \right ).
\end{equation}
Then expectation value at time $t_i = i \Delta t$ is

\begin{equation}
 \E \left (  e^{i u \sigma_i A_i \varphi (t- i \Delta t)} \right )= \E_{\sigma_i}  \left \{ \Phi_{A_i}\left [u \varphi (t- i \Delta t) \right ]   \right \}.
\end{equation}
The probability for there being a jump (i.e., $\sigma=1$) is $\lambda \Delta t$ and the probability of no jump (i.e., $\sigma=0$) is $1-\lambda \Delta t$. Then

\begin{equation}
 \E  \left (  e^{i u \sigma_i A_i \varphi (t- i \Delta t)} \right )= 1 + \lambda \Delta t \left ( \Phi_{A_i}\left [u \varphi (t- i \Delta t) \right ]  -1 \right ) \approx
  \exp \left (- \lambda \Delta t \left \{ 1- \Phi_{A_i}\left [u \varphi (t- i \Delta t) \right ]  \right \} \right ).
\end{equation}
The product over all time steps yields \eqref{Campbell}. If the amplitude outcomes are $\pm A$ with equal probabilities, then 

\begin{equation}
\E \left (  e^{i u X(t)} \right ) = \exp \left \{ - \lambda \int_{0}^{t} d \alpha \left [ 1 - \cos \left ( u A \varphi (t- \alpha) \right ) \right ] \right \}.
\end{equation}

\subsection{Two-time characteristic function}

The two-time characteristic function is

\begin{equation}\label{ch_func_2t}
\Phi_x (u_1,t_1;u_2,t_2) = \E \left (  e^{i \left [ u_1 X(t_1)+ u_2 X(t_2) \right ]} \right ).
\end{equation}
Let us consider the case where $t_2 > t_1$. Then $X(t_1)$ depends on all the jumps before $t_1$ and $X(t_2)$ depends on all jumps before $t_2$, which includes all those that contributed to $X(t_1)$: this is the source of correlation between the two variables. If we split up $X(t_2)$ as

\begin{equation}
  X(t_2) = \sum_{\{k:t_1<t_k<t_2\}} A_k \varphi(t_2-t_k) +\sum_{\{k:t_1>t_k\}} A_k \varphi(t_2-t_k),
\end{equation}
the exponent in Eq.~\eqref{ch_func_2t} is expressed as the sum of independent quantities

\begin{equation}
  u_1X(t_1)+u_2X(t_2) =  \sum_{\{k:t_k<t_1\}} A_k \left[u_1\varphi(t_1-t_k)+u_2\varphi(t_2-t_k)\right] +
  \sum_{\{k:t_1<t_k<t_2\}} u_2 A_k \varphi(t_2-t_k).
\end{equation}
Therefore, the characteristic function becomes

\begin{equation}
\Phi_x (u_1,t_1;u_2,t_2) = \E \left (  e^{i  \sum_{\{k:t_1>t_k\}} A_k \left[u_1\varphi(t_1-t_k)+u_2\varphi(t_2-t_k)\right]} \right )
\E \left (  e^{i \sum_{\{k:t_1<t_k<t_2\}} u_2 A_k \varphi(t_2-t_k)} \right ).
\end{equation}
Following the same procedure we followed to derive the one-time characteristic function, we arrive to

\begin{align}
\nonumber
  \Phi_X(u_1,t_1;u_2,t_2) = & \exp \Bigg  \{ -\lambda\int_{t_1}^{t_2} \left[1-\Phi_A(i u_2\varphi(t_2-\alpha))\right]d\alpha \\ &-\lambda\int_{0}^{t_1}  \left[1-\Phi_A(i u_1\varphi(t_1-\alpha)+ i u_2\varphi(t_2-\alpha))\right]d\alpha \Bigg \}.
\end{align}
If the amplitude outcomes are $\pm A$ with equal probabilities, 

\begin{align}\label{ch_func_2t_final}
\nonumber
  \Phi_X(u_1,t_1;u_2,t_2) = & \exp \Bigg  \{ -\lambda\int_{t_1}^{t_2} \left[1-\cos \left ( u_2\varphi(t_2-\alpha) \right )\right]d\alpha \\ &-\lambda\int_{0}^{t_1}  \left [1-\cos \left ( u_1\varphi(t_1-\alpha)+ u_2\varphi(t_2-\alpha) \right )\right]d\alpha \Bigg \}.
\end{align}
The correlations arise from the second term. For stationary correlations the lower limit should be taken to $-\infty$. It's clear that the coefficients of $u_1 u_2^3$ and $u_2 u_1^3$ will be different, which indicates the absence of time reversal invariance. By Taylor expanding the characteristic function, we have access to all two-time correlators. Specifically,

\begin{align}
\E \left ( X(t_1) X^{3}(t_2)\right ) &= \lambda I_{1,3}^{-\infty, t_1} + 3 \lambda^2 I_{1,1}^{-\infty,t_1} \left ( I_{0,2}^{-\infty,t_1} + I_{0,2}^{t_1,t_2} \right ), \\
\E \left ( X^3(t_1) X(t_2)\right ) &= \lambda I_{3,1}^{-\infty, t_1} + 3 \lambda^2 I_{1,1}^{-\infty,t_1}  I_{2,0}^{-\infty,t_1}, \text{ where} \\
I_{n,m}^{t, t'} &= \int_{t}^{t'} d \alpha \ \varphi^n(t_1-\alpha) \varphi^m(t_2-\alpha).
\end{align}
This result is general for any characteristic function of the form shown in Eq.~\eqref{ch_func_2t_final}, taking the lower limit to $-\infty$.

\end{document}